\documentclass[english,aps,prx,twocolumn,superscriptaddress,longbibliography]{revtex4-2}
\usepackage{amssymb}
\usepackage{amsmath}
\usepackage{graphicx}
\usepackage{natbib}
\usepackage{epstopdf}
\usepackage{color}
\usepackage{braket}
\usepackage{physics}
\usepackage{mathrsfs}
\usepackage{upgreek}

\definecolor{new}{rgb}{.38,.6,.38}
\definecolor{old}{rgb}{1,0,0}

\begin{document}
\title{Probing the Spatial Variation of the Inter-Valley Tunnel Coupling in a Silicon Triple Quantum Dot}

\author{F. Borjans}
\affiliation{Department of Physics, Princeton University, Princeton, New Jersey 08544, USA}
\author{X. Zhang}
\affiliation{Department of Physics, Princeton University, Princeton, New Jersey 08544, USA}
\author{X. Mi}
\altaffiliation{Present address: Google Inc., Santa Barbara, California 93117, USA}
\affiliation{Department of Physics, Princeton University, Princeton, New Jersey 08544, USA}
\author{G. Cheng}
\affiliation{Princeton Institute for Science and Technology of Materials,
Princeton University, Princeton, New Jersey 08544, USA}
\author{N. Yao}
\affiliation{Princeton Institute for Science and Technology of Materials,
Princeton University, Princeton, New Jersey 08544, USA}
\author{C. A. C. Jackson}
\affiliation{HRL Laboratories LLC, 3011 Malibu Canyon Road, Malibu, California 90265, USA}
\author{L. F. Edge}
\affiliation{HRL Laboratories LLC, 3011 Malibu Canyon Road, Malibu, California 90265, USA}
\author{J. R. Petta}
\affiliation{Department of Physics, Princeton University, Princeton, New Jersey 08544, USA}

\begin{abstract}
Electrons confined in silicon quantum dots exhibit orbital, spin, and valley degrees of freedom. The valley degree of freedom originates from the bulk bandstructure of silicon, which has six degenerate electronic minima. 
The degeneracy can be lifted in silicon quantum wells due to strain and electronic confinement, but the ``valley splitting" of the two lowest lying valleys is known to be sensitive to atomic-scale disorder.  Large valley splittings are desirable to have a well-defined spin qubit. In addition, an understanding of the inter-valley tunnel coupling that couples different valleys in adjacent quantum dots is extremely important, as the resulting gaps in the energy level diagram may affect the fidelity of charge and spin transfer protocols in silicon quantum dot arrays. Here we use microwave spectroscopy to probe spatial variations in the valley splitting, and the intra- and inter-valley tunnel couplings ($t_{ij}$ and $t'_{ij}$) that couple dots $i$ and $j$ in a triple quantum dot (TQD). We uncover large spatial variations in the ratio of inter-valley to intra-valley tunnel couplings $t_{12}'/t_{12}=0.90$ and $t_{23}'/t_{23}=0.56$. By tuning the interdot tunnel barrier we also show that $t'_{ij}$
scales linearly with $t_{ij}$, as expected from theory. The results indicate strong interactions between different valley states on neighboring dots, which we attribute to local inhomogeneities in the silicon quantum well. 
\end{abstract}

\maketitle

\section{Introduction}
Continuous research on electron spin qubits defined in silicon quantum dots has led to increasingly impressive levels of quantum control, with recent demonstrations of high single qubit fidelities \cite{Yoneda2018,yang_silicon_2019,andrews_quantifying_2019} and $>$90\% two-qubit gate fidelities \cite{Huang2018,xue_benchmarking_2019}. Progress has been fueled by an investment in high quality Si/SiGe heterostructures \cite{schaffler_high-mobility_1997,deelman_metamorphic_2016}, coupled with the advent of accumulation-mode device designs that are less sensitive to disorder and enable fine control over quantum dot electrons \cite{angus_gate-defined_2007,borselli_undoped_2015,ZajacScalable,lawrie_quantum_2020}. For example, through time-domain control of the quantum dot confinement potential, it is now feasible to shuttle a single charge down an array of nine silicon quantum dots in $\sim$50~ns \cite{Mills2018}.
 
Control of valley states in Si quantum devices, especially those based on Si/SiGe heterostructures, is an oustanding challenge \cite{zwanenburg2013silicon,goswami_controllable_2007,penthorn_two-axis_2019}. While spin-1/2 electrons are often viewed as a canonical two-level system, the valley degree of freedom in the electronic bandstructure of silicon can give rise to low-lying valley-orbit states \cite{friesen_magnetic_2006,friesen_theory_2010,tariq_effects_2019}. For single quantum dots, low-lying valley states inhibit spin initialization and readout routines that are based on energy dependent tunneling \cite{elzerman2004single}. Moreover, spin-valley mixing leads to spin-relaxation hotspots when the Zeeman splitting of the single electron spin state is comparable to the valley splitting \cite{yang2012,petit_spin_2018,Borjans2018,zhang_giant_2020}.

Complications associated with valley splitting are exacerbated in silicon quantum dot arrays. The lifting of the $\pm z$-valley degeneracy that gives rise to valley splitting is set by the abruptness of the electronic interfaces that break inversion symmetry \cite{friesen_valley_2007}. Spatial inhomogeneities in the structure of the quantum well therefore lead to dot-to-dot variations in the valley splitting \cite{culcer_interface_2010,jiang_effects_2012,neyens_critical_2018}. In Si/SiGe systems the valley splitting often lies between $25\rm\,  - 300\rm\, \mu eV$ \cite{borselli_measurement_2011,zajac2015reconfigurable,mi_high-resolution_2017,Borjans2018,ferdous_valley_2018,hollmann_large_2020}. Due to tighter confinement in the $z$-direction, valley splittings $>$200 $\mu$eV have been observed in Si-MOS systems \cite{yang2013spin,gamble_valley_2016,yang_operation_2020}.

The magnitude of the valley splitting is generally sufficient to understand the consequences on individual spins in isolated quantum dots. However, the intra-valley $t_{ij}$ and inter-valley $t'_{ij}$ tunnel couplings will influence the energy level structure of a linear array of tunnel coupled quantum dots. In particular, the location of anti-crossings in the energy level diagram will be set by the valley splitting of each dot $E_{\mathrm{VS},i}$, and the magnitude of the avoided crossings in the energy level diagram will be set by $t_{ij}$ and $t'_{ij}$ \cite{shiau_valley_2007, culcer_quantum_2010,gamble_disorder-induced_2013,russ_theory_2020}. Non-adiabatic transitions, Landau-Zener-Stuckelberg-Majorana interference, and leakage into higher lying energy levels may all influence the fidelity of spin transfer protocols \cite{Zener1932,zhao2018coherent,ginzel_spin_2020}.

Here we investigate valley splitting and inter-valley tunnel coupling using microwave spectroscopy. A triple quantum dot (TQD) is embedded in a superconducting cavity in the circuit quantum electrodynamics device architecture (cQED) \cite{wallraff_strong_2004}. The dipole moment of an electron confined in a quantum dot couples to the electromagnetic field of the superconducting cavity \cite{petersson2012,frey_dipole_2012,viennot_coherent_2015,stock2017,Mi2017}. By probing the microwave transmission through the cavity we sensitively map out the energy level structure of the TQD. To take a first step towards measuring the spatial dependence of valley parameters, we extract the valley splitting of each dot and the intra- and inter-valley tunnel couplings in the TQD. We find significant variations in the valley parameters across the $\sim$200 nm length scale of the device. Valley splittings range from 38 -- 55$\,\rm \mu eV$ and the inter-valley tunnel coupling $t'_{12}$ between dots 1 and 2 is nearly two times larger than $t'_{23}$. These results have important implications on future experiments aimed at demonstrating coherent spin shuttling in quantum dot arrays and reinforce the need for additional improvements in the growth of Si/SiGe heterostructures, especially interface abruptness.

\begin{figure}[tbp]
	\centering
	\includegraphics[width=\columnwidth]{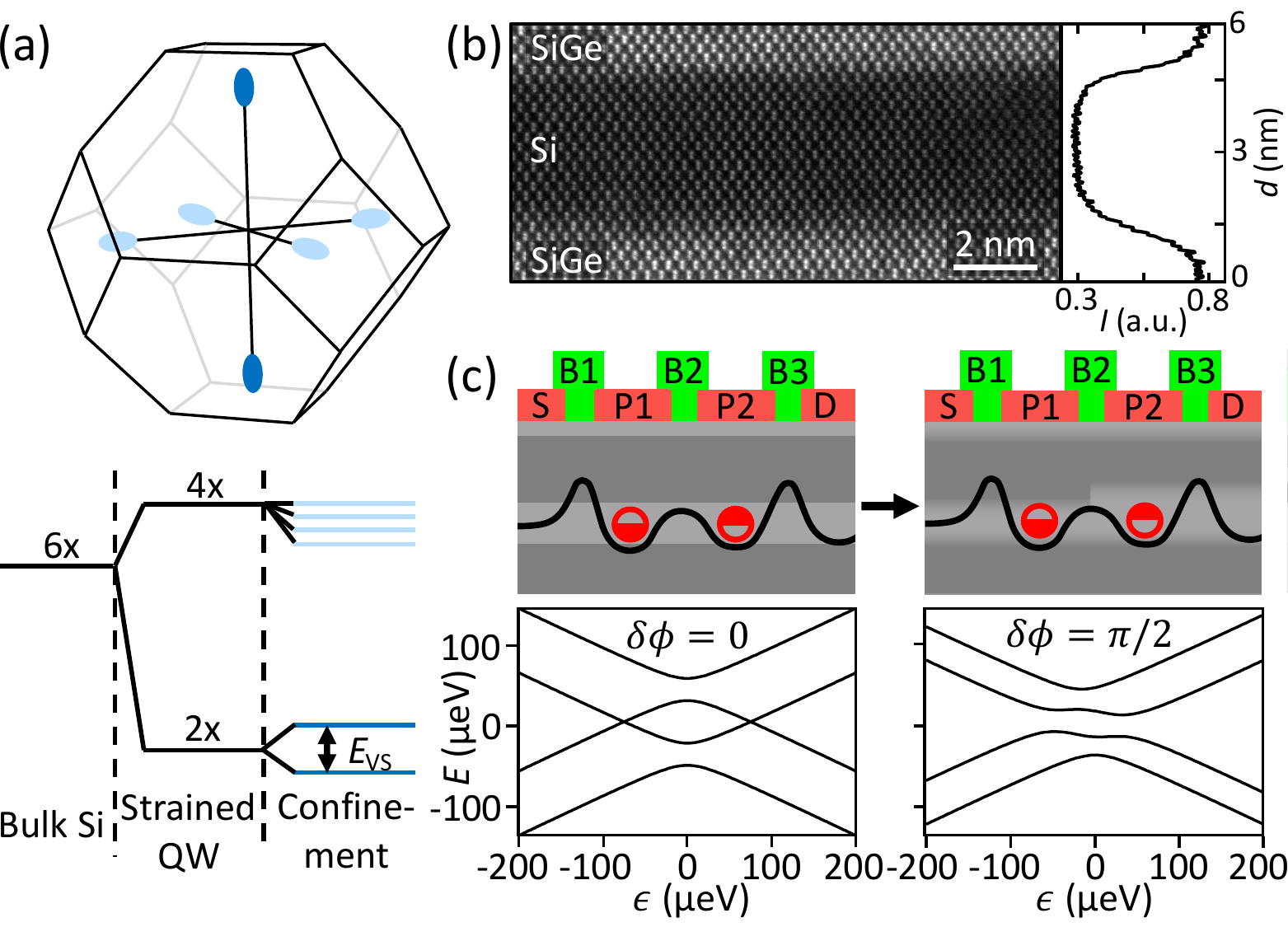} 
	\caption{Valley physics in silicon. (a) First Brillouin zone of silicon, with six degenerate conduction band minima. Tensile strain in the Si quantum well separates the $\pm z$-valleys from the $\pm x,y$-valleys, while vertical confinement lifts the degeneracy of the $\pm z$-valleys. (b) TEM image of the Si/SiGe quantum well. (c) Schematic cross-section of a DQD with corresponding energy levels. An ideal Si/SiGe quantum well with abrupt interfaces (left) leads to large and uniform valley splittings, and no inter-valley tunnel coupling, while a realistic quantum well with soft interfaces and step-edges (right) will have small, non-uniform valley splittings, and strong inter-valley tunnel coupling between the quantum dots. Corresponding energy level diagrams are shown in the bottom graphs.}
	\label{fig:1}
\end{figure}

\begin{figure*}[tbp]
	\centering
	\includegraphics[width=2\columnwidth]{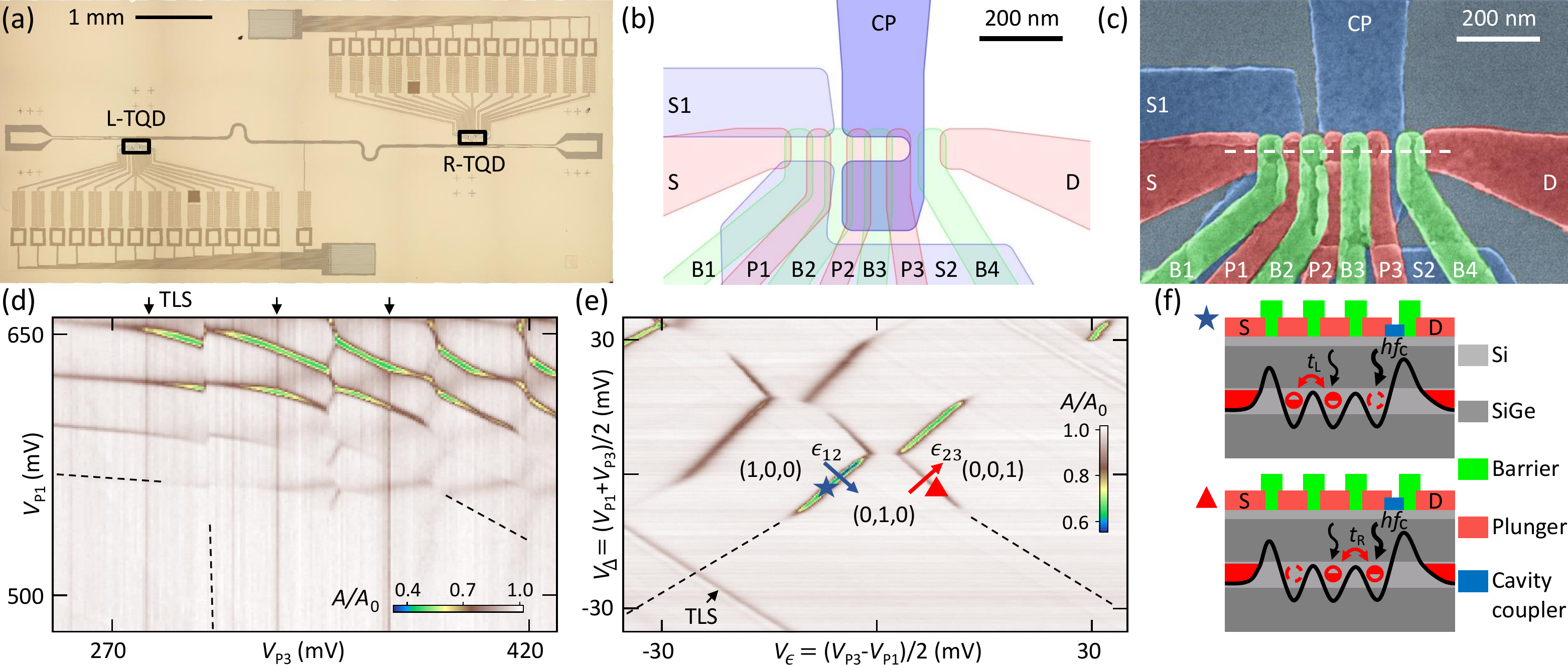}
	\caption{Cavity-coupled TQD. (a) Optical image of a superconducting cavity coupled to two TQDs. (b) Schematic of the TQD device with cavity coupler gate (CP). (c) SEM image of the TQD. (d) Large scale charge stability diagram of the TQD. The dashed black lines indicate the three quantum dot charge transitions. Features corresponding to two level systems (TLS) are indicated by black arrows. (e) Charge stability diagram in the single electron regime. The blue star indicates the (1,0,0)-(0,1,0) interdot transition and the red triangle indicates the (0,1,0)-(0,0,1) interdot transition. (f) Schematic of the operating points indicated by the symbols in (e). Due to the geometry of CP, the microwave coupling to dot 3 is strongest, as indicated by the thickness of the black arrows.
}
	\label{fig:2}
\end{figure*}

\section{Valley states in silicon}
\label{sec:valley}
The band structure of bulk silicon has six degenerate conduction band minima that are located close to the boundary of the first Brillouin zone, as depicted in Fig.\ \ref{fig:1}(a) \cite{ando1982,zwanenburg2013silicon}. Due to the slight difference in lattice constant of the Si quantum well and the surrounding Si$_{0.7}$Ge$_{0.3}$ buffer layers, the Si quantum well is under tensile strain, which increases the energy of the $\pm x,y$-valleys relative to the $\pm z$-valleys \cite{schaffler_high-mobility_1997}. Sharp quantum well interfaces break inversion symmetry and couple the two $\pm z$-valleys, resulting in valley splitting \cite{zwanenburg2013silicon}.

Specifically, the Hamiltonian of the subsystem consisting of the $\pm z$-valleys on dot $i$ can be written in the $\{\ket{i,+z},\ket{i,-z}\}$ basis as
\begin{equation}
H_{V,i} = \begin{pmatrix}
0 & \Delta_i \\
\Delta_i^* & 0
\end{pmatrix},
\end{equation}
with complex valley coupling matrix element $\Delta_i=|\Delta_i|e^{-i\phi_i}$ \cite{culcer_interface_2010,zhao2018coherent}. 
The eigenstates of this system are $\ket{i,\pm}=\tfrac{1}{\sqrt{2}}\left(\ket{i,+z}\pm e^{i\phi_i}\ket{i,-z}\right)$, with eigenenergies $E_{\pm}=\pm |\Delta_i|$. Consequently, the valley splitting of dot $i$ is $E_{\mathrm{VS},i}=E_+-E_-=2|\Delta_i|$ and $\phi_i$ is the valley-orbit phase. The magnitude and phase of $\Delta_i$ are physically rooted in the local properties of the quantum well \cite{culcer_interface_2010,zhao2018coherent}. For example, spatial variations in interface abruptness and disorder will cause $\Delta_i$ to be a function of position in the plane of the quantum well, $\Delta_i$ = $\Delta_i(x,y)$, hence the term ``valley-orbit coupling."

While the valley-orbit phase does not immediately affect the energy levels of a single quantum dot, its consequences are evident when considering a system of two tunnel coupled quantum dots (dots $i$ and $j$). In the basis $\{\ket{i,+z},\ket{i,-z},\ket{j,+z},\ket{j,-z}\}$, the Hamiltonian can be expressed as
\begin{equation}
H_{ij}(\epsilon_{ij})=\begin{pmatrix}
\tfrac{\epsilon_{ij}}{2} & \Delta_i & t_c & 0\\
\Delta_i^* & \tfrac{\epsilon_{ij}}{2} & 0 & t_c\\
t_c & 0 & -\tfrac{\epsilon_{ij}}{2} & \Delta_j\\
0 & t_c & \Delta_j^* & -\tfrac{\epsilon_{ij}}{2}
\end{pmatrix},
\end{equation}
with tunnel coupling $t_c$ between identical $z$-valleys of dots $i$ and $j$, and no tunnel coupling between opposite valleys. The energy level detuning between dots $i$ and $j$ is denoted by $\epsilon_{ij}$.

We diagonalize the local single dot valley dynamics of dots $i$ and $j$ by transforming into the basis $\{\ket{i,+},\ket{i,-},\ket{j,+},\ket{j,-}\}$. We obtain
\begin{equation}
\label{eq:hamiltonian}
H_{ij}'(\epsilon_{ij})=\begin{pmatrix}
\tfrac{\epsilon_{ij}}{2}+E_{\mathrm{VS},i} & 0 & t_{ij} & t_{ij}'\\
0 & \tfrac{\epsilon_{ij}}{2} & t_{ij}' & t_{ij}\\
t_{ij}^* & t_{ij}'^* & -\tfrac{\epsilon_{ij}}{2}+E_{\mathrm{VS},j} & 0\\
t_{ij}'^* & t_{ij}^* & 0 & -\tfrac{\epsilon_{ij}}{2}
\end{pmatrix},
\end{equation}
with intra- and inter-valley tunnel coupling $t_{ij}=\tfrac{1}{2}t_c(1+e^{-i\delta\phi_{ij}})$ and $t_{ij}'=\tfrac{1}{2}t_c(1-e^{-i\delta\phi_{ij}})$, and valley phase difference $\delta\phi_{ij}=\phi_i-\phi_j$ \cite{burkard2016dispersive,zhao2018coherent}. This Hamiltonian results in four energy bands with two intra-valley and two inter-valley anti-crossings determined by $t_{ij}$ and $t_{ij}'$, respectively.

In practice, the valley splitting and phase are strongly dependent on the valley coupling dynamics at the Si/SiGe quantum well interfaces. Figure \ref{fig:1}(b) shows a high-resolution transmission electron microscope (TEM) image of the heterostructure. To the right, the row-averaged intensity of the image is plotted. The interface is of fairly high quality with an overall roughness of only a few lattice sites over the displayed range. However, even a small number of atomic steps in the interface can have a large impact on the valley phase difference and overall structure of the DQD energy level diagram \cite{friesen_magnetic_2006,friesen_valley_2007,shiau_valley_2007,gamble_disorder-induced_2013}. A uniform (step-edge-free) interface in the region of the DQD corresponds to the limit $\delta\phi_{ij}$ = 0, which results in a set of bonding-antibonding charge states for each of the valleys (see Fig.\ 1(c), left panel). Realistic quantum wells will have less abrupt interfaces and step edges that result in intervalley tunnel coupling (see Fig.\ 1(c), right panel). For example, with $\delta\phi_{ij}$ = $\pi/2$, there are four avoided crossings in the energy level diagram, indicating the presence of both valley conserving and non-conserving charge transitions between dots $i$ and $j$. Moreover, smooth, disordered interfaces will generally yield smaller $E_{\mathrm{VS}}$ that exhibit dot-to-dot variations \cite{culcer_interface_2010}. Comprehensive studies of the full set of valley-parameters have so far not been conducted. Microwave spectroscopy of a TQD in the cQED architecture provides an opportunity to sensitively probe the spatial variation of valley states in a silicon quantum device.

\section{Experimental setup}

In this experiment a half-wavelength $\lambda/2$ superconducting cavity with resonance frequency $f_c=6.76\,\rm GHz$ and photon loss rate $\kappa/2\pi = 1.4\,\rm MHz$ is coupled to two Si/SiGe TQDs [Fig.\ \ref{fig:2}(a)]. The TQDs are fabricated using an overlapping gate architecture \cite{zajac2015reconfigurable}. The center pin of the cavity is galvanically connected to the cavity-coupler gate ``CP," as shown schematically in Fig.\ \ref{fig:2}(b). Along with screening gates S1 and S2, the CP gate is part of the first of three overlapping Al gate layers \cite{Borjans_APL_2020}. Plunger (barrier) gates are defined in the second (third) aluminum layers and the layers are electrically isolated from one another by a native Al$_2$O$_3$ oxide barrier.

The TQD chemical potentials are controlled with plunger gates P1--P3, while the interdot barriers and barriers to the source (S) and drain (D) reservoirs are controlled by gates B1--B4. To enhance the dot-cavity coupling, the CP gate is designed to wrap around dot 3, which effectively adds a potential barrier between dot 3 and the D reservoir, in addition to the barrier defined by gate B4.  A scanning electron microscope (SEM) image of the TQD is shown in Fig.\ \ref{fig:2}(c), with a white dashed line indicating the long-axis of the TQD. Details related to the fabrication of similar devices have been presented elsewhere \cite{MiAPL2017}. The data sets in this manuscript are acquired from the TQD located at the right anti-node of the cavity (denoted R-TQD).

We map out the TQD charge stability diagram [Fig.\ \ref{fig:2}(d)] by measuring the normalized cavity transmission $A/A_0$ as a function of the gate voltages $V_{\mathrm{P1}}$ and $V_{\mathrm{P3}}$ \cite{petersson2012}. When sweeping $V_{\mathrm{P1}}$ and $V_{\mathrm{P3}}$ we observe nearly horizontal charge transitions in the cavity response that correspond to charge transfer between dot 1 and the S reservoir. As a result of the mutual charging energy, the dot 1 charge transitions shift abruptly when an electron is added or removed from dot 3 (see vertical dashed line). As the CP gate limits the tunneling rate to the D reservoir in the few electron regime, no direct cavity response is observed for dot 3 charge transitions. The absence of dot 1 transition shifts for $V_{\mathrm{P3}}$ below the indicated dashed line however allows us to verify the single electron occupation of dot 3. Higher electron occupancies would otherwise affect the valley-dynamics \cite{corrigan_coherent_2020}. Due to cross-coupling of the gate voltages $V_{\mathrm{P1}}$ and $V_{\mathrm{P3}}$ to dot 2, we observe angled dot 2 charge transitions in the data, which interact with the dot 1 and 3 charging transitions.

For the remainder of the manuscript we operate in the single electron regime. By performing pairwise sweeps of $V_{\mathrm{P1}}$/$V_{\mathrm{P2}}$ and $V_{\mathrm{P2}}$/$V_{\mathrm{P3}}$, we tune the device towards the (1,0,0)-(0,1,0)-(0,0,1) transition shown in Fig.\ \ref{fig:2}(e). Here we use the virtual gates $V_\epsilon=(V_{\mathrm{P3}}-V_{\mathrm{P1}})/2$ and $V_{\Delta}=(V_{\mathrm{P1}}+V_{\mathrm{P3}})/2$ to access the different charge states of the TQD. In this tuning configuration, the tunneling rates to the S and D reservoirs are suppressed relative to $f_c$, and the primary cavity response stems from the (1,0,0)-(0,1,0) and (0,1,0)-(0,0,1) interdot charge transitions. We denote these transitions with a blue star and a red triangle, respectively, and show a schematic of the device cross-section in Fig.\ \ref{fig:2}(f). The cross-section corresponds to the device area indicated by the white dashed line in Fig.\ \ref{fig:2}(c). The asymmetric geometry of the CP gate creates an electric field gradient across all three dots, allowing us to observe a signal at both interdot charge transitions.

\section{Cavity response to valley states}

\begin{figure}[tbp]
	\centering
	\includegraphics[width=\columnwidth]{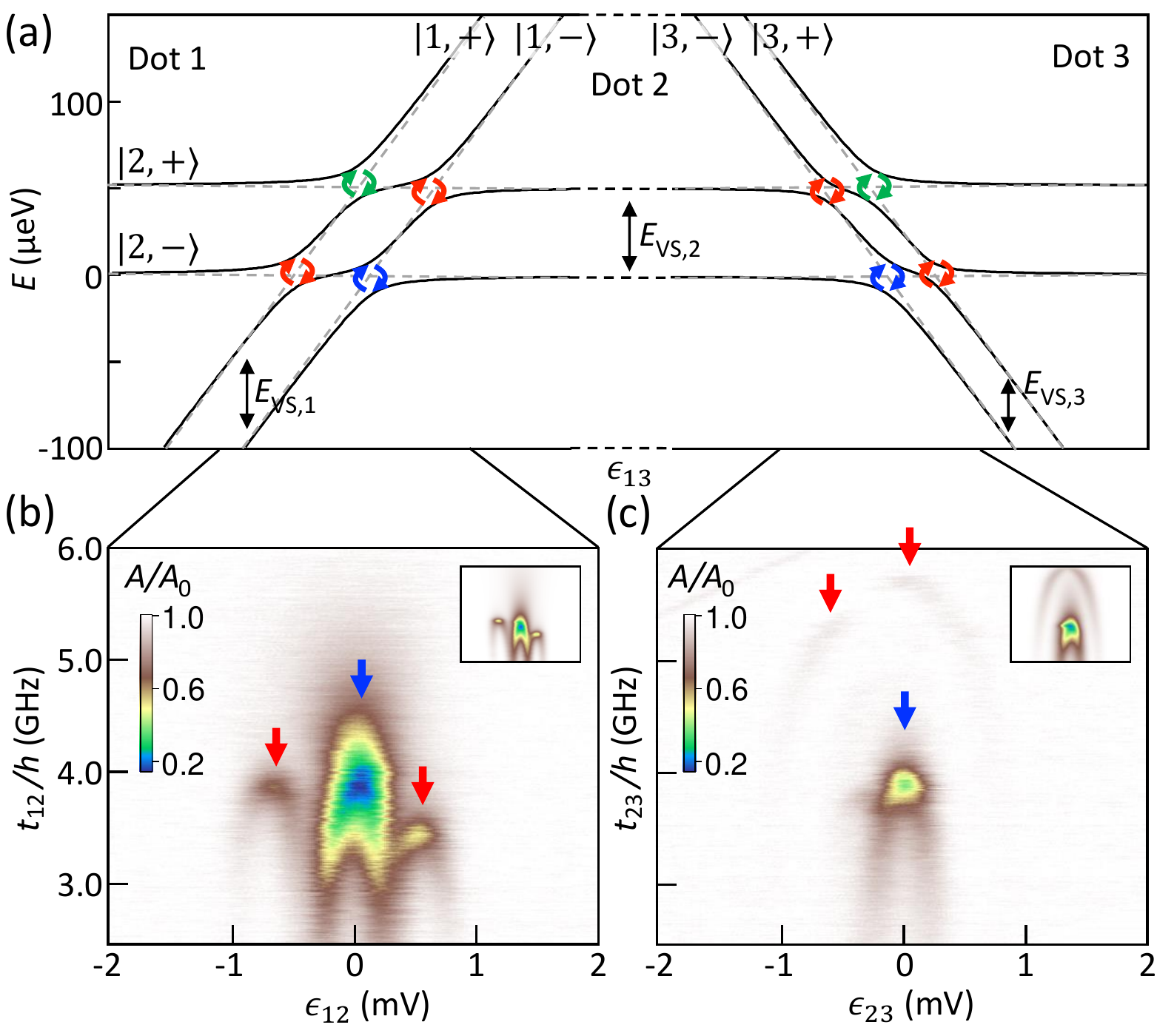}
	\caption{Cavity response in the single electron regime at $T_e=350\rm\,mK$. (a) Energy bands of the TQD, with valley splittings denoted by $E_{\mathrm{VS},i}$. At low temperatures, only the transition between the lowest energy states (blue) is visible in the cavity response. By increasing temperature, inter-valley transitions (red) can be accessed. (b) Cavity response $A/A_0$ near the (1,0,0)-(0,1,0) transition as a function of $t_{12}$ and the detuning $\epsilon_{12}$ between dots 1 and 2. The colored arrows correspond to the transitions highlighted in (a). (c) $A/A_0$ in the vicinity of the (0,1,0)-(0,0,1) transition as a function of $t_{23}$ and $\epsilon_{23}$. Insets in (b) and (c) show the simulated cavity response.}
	\label{fig:3}
\end{figure}

To gather insight into the TQD valley physics we sketch the low-lying TQD energy levels as a function of the detuning  $\epsilon_{13}$ between dots 1 and 3 in Fig.\ \ref{fig:3}(a). In this regime the (1,0,0)-(0,1,0) and (0,1,0)-(0,0,1) interdot charge transitions can be analyzed independently. Each interdot transition involves four energy levels. Focusing on the (1,0,0)-(0,1,0) transition the four relevant levels are $\ket{1,\pm}$ and  $\ket{2,\pm}$.

The microwave cavity is sensitive to charge dynamics within the TQD due to dipole coupling between the electron trapped in the device and the cavity electric field \cite{petersson2012,frey_dipole_2012}. In general, the cavity response $A/A_0$ is strongest near charge avoided-crossings in the energy level diagram [see blue, red and green arrows in Fig.\ 3(a)]. At low temperatures where $k_BT_e \ll E_{\mathrm{VS},i}$, with Boltzmann constant $k_B$ and electron temperature $T_e$, the electron primarily resides in the ground state and the excited valley states are not prominent in the cavity response. Participation of the excited valley states $\ket{1,+}, \ket{2,+}, \ket{3,+}$ in the cavity response can be increased by raising the temperature and thermally exciting the charge according to the Boltzmann distribution with occupation probability $p_i = e^{-E_{i}/k_{B}T_{e}}/\sum_{i} e^{-E_{i}/k_{B}T_{e}}$, with state energy $E_{i}$ \cite{burkard2016dispersive,mi_high-resolution_2017}. Increasing the electron temperature, such that $k_BT_e \approx E_{\mathrm{VS},i}$, is effective at increasing the visibility of the higher-lying inter-valley transitions in the cavity response (red transitions). In practice, the excited intra-valley transitions (green transitions) are still masked by the larger participation of the ground state intra-valley transitions at nearly the same voltage. Nevertheless, the cavity response data are sufficient to extract the valley splittings, and intra- and inter-valley tunnel couplings.

Figure \ref{fig:3}(b) displays the cavity transmission $A/A_0$ as a function of the interdot detuning $\epsilon_{12}$ and tunnel coupling $|t_{12}|$ with $T_e=350\,\rm mK$, which is achieved by actively heating the mixing chamber plate of the dilution refrigerator. Here the detuning is plotted in units of mV, as the lever arm varies slightly as a function of $t_{12}$. The ground state transition $\ket{1,-}\leftrightarrow\ket{2,-}$ corresponds to the center arch that is most pronounced in the data. With $\epsilon_{12} \approx 0$, the cavity response is the strongest around $t_{12}/h=4\,\rm GHz$ ($h$ is Planck's constant), where the energy difference of the coupled states is most insensitive to noise on the detuning axis $\epsilon_{12}$ and the transition energy is resonant with the cavity $E_2-E_1=hf_c$. In contrast to the valley-free case \cite{Mi2017, Borjans_APL_2020}, here $E_2-E_1 < 2t_{12}$, as level repulsion from the excited valley states $\ket{1,+}$ and $\ket{2,+}$ reduce its effective energy splitting. 

Next to the strong central feature we observe two similar, but fainter, arches. These arches correspond to the thermally occupied inter-valley transitions $\ket{1,\pm}\leftrightarrow\ket{2,\mp}$ (red arrows). We note that tuning $t_{12}$ by adjusting $V_{B2}$ affects both the inter- and intra-valley signatures in the data. Moreover, the position of the side arches can be qualitatively understood by looking at the energy diagram in Fig. \ref{fig:3}(a). Raising $E_{\mathrm{VS},1}$ will move the $\ket{1,+}\leftrightarrow\ket{2,-}$ transition towards more negative detuning. Similarly, $E_{\mathrm{VS},2}$ affects the horizontal positioning of the right side arch via a shift of the $\ket{1,-}\leftrightarrow\ket{2,+}$ transition. 
From this qualitative analysis and the larger spacing between the left and center feature, we deduce $E_{\mathrm{VS},1}>E_{\mathrm{VS},2}$. To understand the vertical positioning of the arches, we focus on the strongest cavity signal at the maxima of the arches indicated by the blue (red) arrows. At these points the intra(inter)-valley transition energies are dominated by the contribution of the corresponding couplings $t_{12}(t'_{12})$ and coincide with $hf_c$. The observation that these points are reached at different values of $t_{12}$ for the two side arches indicates that in fact $t'_{12}$ changes with detuning in our system, i.e. $t'_{12}=t'_{12}(\epsilon)$. 
This implies that also $t_{12}=t_{12}(\epsilon)$ due to their functional interdependence. Specifically, with $t_{12}(\epsilon=0)/h=3.5\,\rm GHz$, $t'$ is large enough for the $\ket{1,-}\leftrightarrow\ket{2,+}$ transition to be resonant at positive detuning, while it is too small at negative detuning leaving the $\ket{1,+}\leftrightarrow\ket{2,-}$ transition below resonance. Unless the detuning dependence is explicitly specified, we identify $t_{ij}=t_{ij}(\epsilon=0)$ and $t_{ij}'=t_{ij}'(\epsilon=0)$.

For the (0,1,0)-(0,0,1) transition [Fig.\ 3(c)], we observe similar behavior for the $\ket{2,-}\leftrightarrow\ket{3,-}$ intra-valley transition. However, the $\ket{2,\pm}\leftrightarrow\ket{3,\mp}$ inter-valley features do not appear in the shape of two arches next to the main feature, but they merge into one bigger arch at much higher $t_{23}$. The qualitative differences between the datasets indicates a variation of the valley parameters across the TQD. The asymmetries caused by the variation of $t$ and $t'$ can be reproduced by input-output simulations shown in the insets of Figs. \ref{fig:3}(b,c).

\section{Quantitive Extraction of the Valley Parameters}

\begin{figure}[tbp]
	\centering
	\includegraphics[width=\columnwidth]{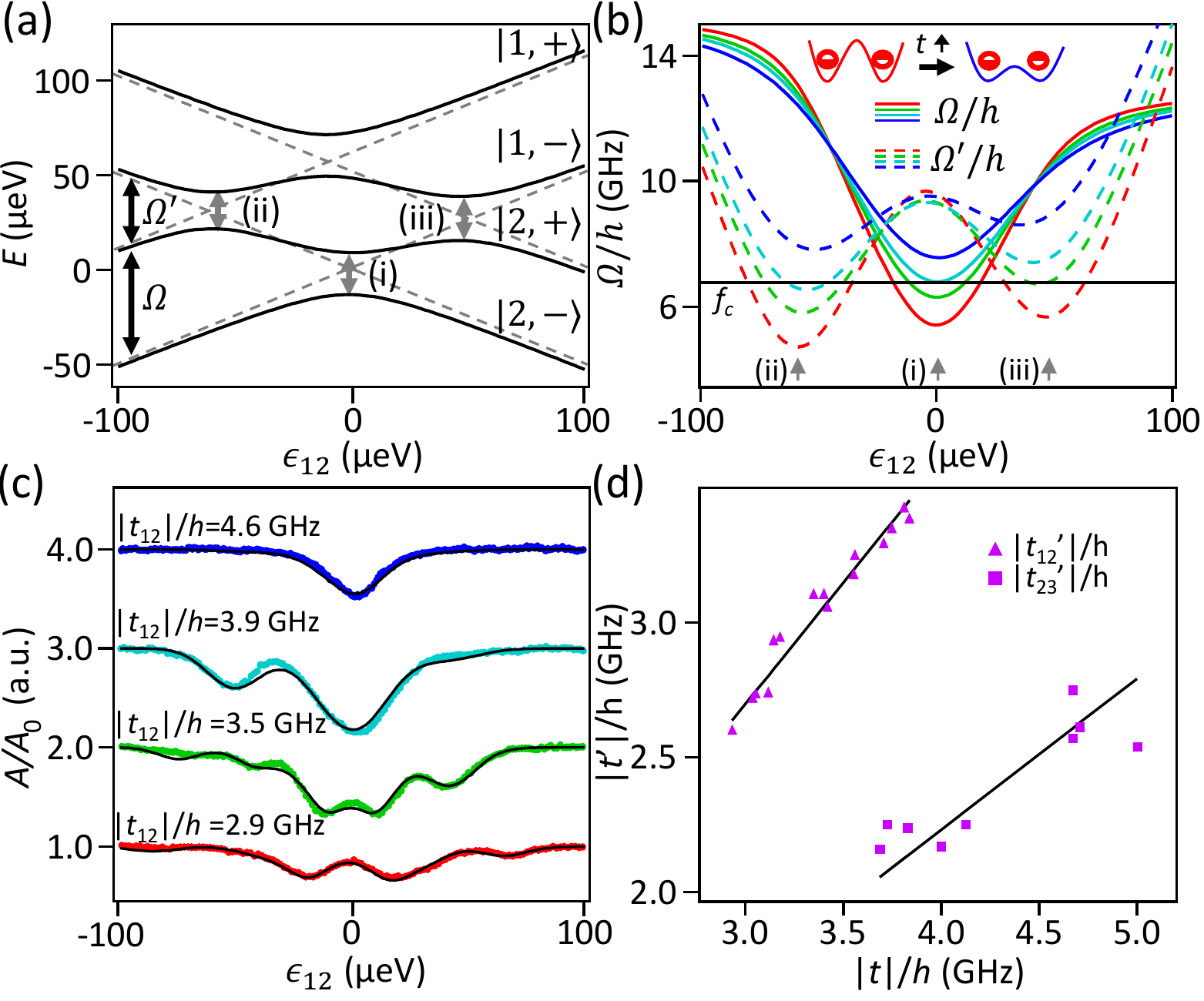}  
	\caption{Quantitative analysis of the cavity response. 
	(a) Energy level diagram for $|t_{12}|/h=2.9\rm\,GHz$. Transitions $\Omega$ and $\Omega^\prime$ are energetically accessible at $T_e=350\rm\,mK$. (b) Transition frequencies $\Omega/h$ and $\Omega^\prime/h$ as a function of $\epsilon_{12}$ for the four values of $t_{12}$ indicated in (c). The minima in the transition frequencies are labeled (i)-(iii). (c) Line cuts extracted from Fig.\ \ref{fig:3}(b) for different $|t|$ (fits are shown in black). The line cuts correspond to the low $|t|$ regime, resonant regimes for both intra and inter-valley transitions, and the high $|t|$ regime. (d) Extracted inter-valley tunnel coupling rates $|t’_{ij}|/h$ as a function of $|t_{ij}|/h$ for both interdot transitions.}
	\label{fig:4}
\end{figure}

We now quantitatively analyze the cavity response data in light of the theoretical background given in section \ref{sec:valley}. The energy levels in the vicinity of the (1,0,0)-(0,1,0) transition are plotted in Fig.\ \ref{fig:4}(a). The dashed lines indicate the expected energy levels with no tunnel coupling, while the solid lines show the energy levels with $|t_{12}|/h=2.9\,\rm GHz$ and $|t_{12}'|/h=2.6\,\rm GHz$. We denote the intra-valley transition $\ket{1,-}\leftrightarrow\ket{2,-}$ as (i), and the inter-valley transitions $\ket{1,\pm}\leftrightarrow\ket{2,\mp}$ as (ii) and (iii), respectively. We plot the intra- and inter-valley transition frequencies $\Omega/h$ and $\Omega^\prime/h$ as a function of $\epsilon_{12}$ in Fig.\ \ref{fig:4}(b), and the cavity frequency $f_c$ is shown for comparison. An appreciable cavity response is observed near detunings $\epsilon_{12}$ where $\Omega/h$ or $\Omega^\prime/h$ $\approx$ $f_c$. $\Omega/h$ exhibit a single minimum corresponding to the ground state intra-valley transition (i) close to $\epsilon=0$. Similarly, the two inter-valley transitions (ii) and (iii) lead to two minima in $\Omega^\prime/h$ at finite detuning.

We next utilize Figs.\ \ref{fig:4}(a,b) to identify the features in the cavity response data. Four linecuts extracted from the Fig.\ \ref{fig:3}(b) data set are plotted in Fig.\ \ref{fig:4}(c), and show the evolution of $A/A_0$ as a function of $\epsilon_{12}$ as $t_{12}$ is reduced. Qualitatively, the complexity of the cavity response increases as $|t_{12}|/h$ is reduced from $4.6\,\rm GHz$ to $2.9\,\rm GHz$. 
At $|t_{12}|/h=4.6\,\rm GHz$, both transition energies lie above resonance for the whole range of $\epsilon_{12}$, with $\Omega/h$ only approaching $f_c$ at $\epsilon_{12}=0$. At this point, the inter-valley features are very weak and only the intra-valley transition (i) contributes to an appreciable dispersive signal in the cavity transmission.
At $|t_{12}|/h=3.9\,\rm GHz$, both (i) and (ii) are nearly resonant with the cavity, leading to a strong suppression of $A/A_0$ at $\epsilon_{12}=0$, and a local minimum at negative $\epsilon_{12}$. 
Reducing the tunnel coupling further to $|t_{12}|/h=3.5\,\rm GHz$, (iii) is nearly resonant with $f_c$, causing a pronounced reduction of the cavity transmission at positive detuning. As (i) is below resonance, we observe two additional minima in $A/A_0$, corresponding to the two detuning values where $\Omega/h=f_c$.
Finally, when $|t_{12}|/h=2.9\,\rm GHz$, the ground state intra-valley transition (i) and both inter-valley transitions (ii) and (iii) lie below resonance, so that there are in theory six detuning values at which either $\Omega/h$ or $\Omega^\prime/h=f_c$. However, the central four resonance conditions lie pair-wise close to each other, such that their corresponding features merge into two central transmission minima [see lower portion of Fig.\ 3(b), and Fig.\ 4(c)].

To determine the TQD valley parameters we quantitatively analyze linecuts similar to those shown in Fig.\ \ref{fig:4}(c). Superimposed on the data are best fits obtained by numerically diagonalizing the Hamiltonian in Eq.\ \ref{eq:hamiltonian} for each value of $\epsilon_{12}$ and feeding the resulting energies into cavity input-output theory \cite{benito2017input}. To generate the level diagram, we treat $t_{12}=|t_{12}|$ and $t_{12}'=|t_{12}'|$ as real valued parameters for the fits, which leaves the energy levels unaffected \cite{burkard2016dispersive,zhao2018coherent}. To account for the observed asymmetry in the Fig.\ 3(b) inter-valley tunnel couplings at positive and negative detuning, we parameterize a small linear dependence of the intra-valley coupling on detuning $t_{ij}(\epsilon_{ij})=t_{ij}(\epsilon_{ij}=0)+a_{ij}\epsilon_{ij}$. Fit parameters of the Hamiltonian are $E_{\mathrm{VS},i}$, $t_{ij}(\epsilon_{ij}=0)$, $a_{ij}$, leverarm $\alpha_{ij}$ and the valley phase difference $\delta \phi_{ij}$, which is defined by $|t_{ij}'/t_{ij}| = \tan(\delta\phi_{ij}/2)$. We take into account the finite population of the excited states at $T_e=350\,\rm mK$ via the Boltzmann distribution. In addition, we fit the charge-cavity coupling rate $g_{ij}/2\pi$. The inhomogenous broadening of the features caused by charge noise is implemented by a Gaussian convolution of standard deviation $\sigma_{ij}$. Combining the results from the linecut fits we find $g_{12}/2\pi=37\,\rm MHz$ and $\sigma_{12}=9\,\rm\mu eV$. The charge dephasing rate $\gamma_{12}/2\pi=32\,\rm MHz$ is extracted in a separate measurement by probing the ESR linewidth. Similar data (not shown) are acquired near the (0,1,0)-(0,0,1) transition, with best fit parameters $g_{23}/2\pi = 23\,\rm MHz$, $\gamma_{23}/2\pi=39\,\rm MHz$ and $\sigma_{23}=7\,\rm\mu eV$.

We first extract the valley splitting of each quantum dot. Analyzing the linecuts yields $E_{\mathrm{VS},1}=63\,\rm\mu eV$ and $E_{\mathrm{VS},2}$ = $53\,\rm\mu eV$, whereas the data from the (0,1,0)-(0,0,1) transition yields $E_{\mathrm{VS},2}$ = $50\,\rm\mu eV$ and $E_{\mathrm{VS},3}=38\,\rm\mu eV$. The magnitude of the valley splitting is consistent with previous measurements on similar devices \cite{mi_high-resolution_2017}. Crucially, the values of $E_{\mathrm{VS},2}$ obtained from the two independent measurements are self-consistent and imply that the valley splittings are insensitive to small variations in the plunger gate voltages.

Next, we extract values of $|t_{12}'|$ for multiple values of $|t_{12}|$, as shown in Fig.\ \ref{fig:4}(d). We expect a linear relationship between $|t_{ij}'|$ and $|t_{ij}|$ described by the formula $|t_{ij}'/t_{ij}| = \tan(\delta\phi_{ij}/2)$. We determine the scaling relation to be $|t_{12}'/t_{12}|=0.90$ for the (1,0,0)-(0,1,0) transition with $a_{12}=0.02$. Similarly, we extract $|t_{23}'/t_{23}|=0.56$ for the (0,1,0)-(0,0,1) transition with $a_{23}=0$. Using these scaling relationships, we find valley phase differences $\delta\phi_{12}=84^\circ$ and $\delta\phi_{23}=58^\circ$. The substantial valley phase differences point toward a level diagram closely resembling that shown in Fig.\ 4(a), with the intra- and inter-valley tunnel couplings leading to four avoided crossings. In contrast, an abrupt and uniform Si/SiGe interface would yield $\delta\phi_{12}$=0, with only non-zero intra-valley couplings [left panel of Fig.\ 1(c)]. 

\section{Conclusion and outlook}

In conclusion, we have used microwave spectroscopy to probe the energy level structure of a Si TQD containing a single electron. Consistent with previous work, we find significant variations in the valley splitting across the device \cite{borselli_measurement_2011,zajac2015reconfigurable,mi_high-resolution_2017,Borjans2018,ferdous_valley_2018,hollmann_large_2020}. Going beyond previous work, we capitalize on the sensitivity of cQED microwave spectroscopy to the curvature of the quantum dot energy levels to probe spatial variations in the inter-valley tunnel coupling. Consistent wth theory, we find that the inter-valley coupling scales linearly with the intra-valley coupling. Moreover, examination of both the (1,0,0)-(0,1,0) and (0,1,0)-(0,0,1) interdot charge transitions allows us to probe spatial variations in the valley orbit phase. Significant variations in the valley phase difference $\delta \phi_{ij}$ over the small $\sim$200 nm length-scale of the device highlights the importance of improving the Si/SiGe interface quality. 

Our results indicate a significant variation of the valley splitting and valley-orbit phase across the TQD. Variable valley splittings can impact single qubit operation and readout, as rapid spin-valley mixing occurs when the Zeeman energy and valley splitting are comparable, resulting in vastly reduced spin liftetimes \cite{yang2013spin}. Small valley splittings also limit the use of Pauli spin blockade for singlet-triplet readout in silicon \cite{ maune2012}.

Considering efforts to transport spins down large quantum dot arrays \cite{Mills2018}, spatial variations in the valley phase difference $\delta\phi_{ij}$ could be especially problematic. From the Landau-Zener model, the non-adiabatic transition probability at an avoided crossing is exponentially dependent on the square of the coupling matrix element. Ideally, the inter-valley gaps in the level diagram would be zero or small, limiting mixing into excited valley states during spin shuttling. The non-zero valley phase difference, and its variation across the device, could very well impact the performance of spin shuttling protocols. Looking forward, the measurements presented here can be used to provide valuable feedback on the quality of interfaces in the Si/SiGe system.

\begin{acknowledgements}
We thank Guido Burkard and Mark Friesen for critical comments on the manuscript, and Michael Gullans and Stefan Putz for technical contributions to the research. Supported by Army Research Office grant W911NF-15-1-0149 and the Gordon and Betty Moore Foundation's EPiQS Initiative through grant GBMF4535. Devices were fabricated in the Princeton University Quantum Device Nanofabrication Laboratory. The authors acknowledge the use of Princeton’s Imaging and Analysis Center, which is partially supported by the Princeton Center for Complex Materials, a National Science Foundation MRSEC program (DMR-1420541).
\end{acknowledgements}


\bibliography{References_Borjans_PRXQ_2020_v3}

\begin{thebibliography}{55}%
\makeatletter
\providecommand \@ifxundefined [1]{%
 \@ifx{#1\undefined}
}%
\providecommand \@ifnum [1]{%
 \ifnum #1\expandafter \@firstoftwo
 \else \expandafter \@secondoftwo
 \fi
}%
\providecommand \@ifx [1]{%
 \ifx #1\expandafter \@firstoftwo
 \else \expandafter \@secondoftwo
 \fi
}%
\providecommand \natexlab [1]{#1}%
\providecommand \enquote  [1]{``#1''}%
\providecommand \bibnamefont  [1]{#1}%
\providecommand \bibfnamefont [1]{#1}%
\providecommand \citenamefont [1]{#1}%
\providecommand \href@noop [0]{\@secondoftwo}%
\providecommand \href [0]{\begingroup \@sanitize@url \@href}%
\providecommand \@href[1]{\@@startlink{#1}\@@href}%
\providecommand \@@href[1]{\endgroup#1\@@endlink}%
\providecommand \@sanitize@url [0]{\catcode `\\12\catcode `\$12\catcode
  `\&12\catcode `\#12\catcode `\^12\catcode `\_12\catcode `\%12\relax}%
\providecommand \@@startlink[1]{}%
\providecommand \@@endlink[0]{}%
\providecommand \url  [0]{\begingroup\@sanitize@url \@url }%
\providecommand \@url [1]{\endgroup\@href {#1}{\urlprefix }}%
\providecommand \urlprefix  [0]{URL }%
\providecommand \Eprint [0]{\href }%
\providecommand \doibase [0]{https://doi.org/}%
\providecommand \selectlanguage [0]{\@gobble}%
\providecommand \bibinfo  [0]{\@secondoftwo}%
\providecommand \bibfield  [0]{\@secondoftwo}%
\providecommand \translation [1]{[#1]}%
\providecommand \BibitemOpen [0]{}%
\providecommand \bibitemStop [0]{}%
\providecommand \bibitemNoStop [0]{.\EOS\space}%
\providecommand \EOS [0]{\spacefactor3000\relax}%
\providecommand \BibitemShut  [1]{\csname bibitem#1\endcsname}%
\let\auto@bib@innerbib\@empty
\bibitem [{\citenamefont {Yoneda}\ \emph {et~al.}(2018)\citenamefont {Yoneda},
  \citenamefont {Takeda}, \citenamefont {Otsuka}, \citenamefont {Nakajima},
  \citenamefont {Delbecq}, \citenamefont {Allison}, \citenamefont {Honda},
  \citenamefont {Kodera}, \citenamefont {Oda}, \citenamefont {Hoshi},
  \citenamefont {Usami}, \citenamefont {Itoh},\ and\ \citenamefont
  {Tarucha}}]{Yoneda2018}%
  \BibitemOpen
  \bibfield  {author} {\bibinfo {author} {\bibfnamefont {J.}~\bibnamefont
  {Yoneda}}, \bibinfo {author} {\bibfnamefont {K.}~\bibnamefont {Takeda}},
  \bibinfo {author} {\bibfnamefont {T.}~\bibnamefont {Otsuka}}, \bibinfo
  {author} {\bibfnamefont {T.}~\bibnamefont {Nakajima}}, \bibinfo {author}
  {\bibfnamefont {M.~R.}\ \bibnamefont {Delbecq}}, \bibinfo {author}
  {\bibfnamefont {G.}~\bibnamefont {Allison}}, \bibinfo {author} {\bibfnamefont
  {T.}~\bibnamefont {Honda}}, \bibinfo {author} {\bibfnamefont
  {T.}~\bibnamefont {Kodera}}, \bibinfo {author} {\bibfnamefont
  {S.}~\bibnamefont {Oda}}, \bibinfo {author} {\bibfnamefont {Y.}~\bibnamefont
  {Hoshi}}, \bibinfo {author} {\bibfnamefont {N.}~\bibnamefont {Usami}},
  \bibinfo {author} {\bibfnamefont {K.~M.}\ \bibnamefont {Itoh}},\ and\
  \bibinfo {author} {\bibfnamefont {S.}~\bibnamefont {Tarucha}},\ }\bibfield
  {title} {\bibinfo {title} {A quantum-dot spin qubit with coherence limited by
  charge noise and fidelity higher than 99.9\%},\ }\href
  {https://doi.org/10.1038/s41565-017-0014-x} {\bibfield  {journal} {\bibinfo
  {journal} {Nat. Nanotechnol.}\ }\textbf {\bibinfo {volume} {13}},\ \bibinfo
  {pages} {102} (\bibinfo {year} {2018})}\BibitemShut {NoStop}%
\bibitem [{\citenamefont {Yang}\ \emph {et~al.}(2019)\citenamefont {Yang},
  \citenamefont {Chan}, \citenamefont {Harper}, \citenamefont {Huang},
  \citenamefont {Evans}, \citenamefont {Hwang}, \citenamefont {Hensen},
  \citenamefont {Laucht}, \citenamefont {Tanttu}, \citenamefont {Hudson},
  \citenamefont {Flammia}, \citenamefont {Itoh}, \citenamefont {Morello},
  \citenamefont {Bartlett},\ and\ \citenamefont {Dzurak}}]{yang_silicon_2019}%
  \BibitemOpen
  \bibfield  {author} {\bibinfo {author} {\bibfnamefont {C.~H.}\ \bibnamefont
  {Yang}}, \bibinfo {author} {\bibfnamefont {K.~W.}\ \bibnamefont {Chan}},
  \bibinfo {author} {\bibfnamefont {R.}~\bibnamefont {Harper}}, \bibinfo
  {author} {\bibfnamefont {W.}~\bibnamefont {Huang}}, \bibinfo {author}
  {\bibfnamefont {T.}~\bibnamefont {Evans}}, \bibinfo {author} {\bibfnamefont
  {J.~C.~C.}\ \bibnamefont {Hwang}}, \bibinfo {author} {\bibfnamefont
  {B.}~\bibnamefont {Hensen}}, \bibinfo {author} {\bibfnamefont
  {A.}~\bibnamefont {Laucht}}, \bibinfo {author} {\bibfnamefont
  {T.}~\bibnamefont {Tanttu}}, \bibinfo {author} {\bibfnamefont {F.~E.}\
  \bibnamefont {Hudson}}, \bibinfo {author} {\bibfnamefont {S.~T.}\
  \bibnamefont {Flammia}}, \bibinfo {author} {\bibfnamefont {K.~M.}\
  \bibnamefont {Itoh}}, \bibinfo {author} {\bibfnamefont {A.}~\bibnamefont
  {Morello}}, \bibinfo {author} {\bibfnamefont {S.~D.}\ \bibnamefont
  {Bartlett}},\ and\ \bibinfo {author} {\bibfnamefont {A.~S.}\ \bibnamefont
  {Dzurak}},\ }\bibfield  {title} {\bibinfo {title} {Silicon qubit fidelities
  approaching incoherent noise limits via pulse engineering},\ }\href
  {https://doi.org/10.1038/s41928-019-0234-1} {\bibfield  {journal} {\bibinfo
  {journal} {Nat. Electron.}\ }\textbf {\bibinfo {volume} {2}},\ \bibinfo
  {pages} {151} (\bibinfo {year} {2019})}\BibitemShut {NoStop}%
\bibitem [{\citenamefont {Andrews}\ \emph {et~al.}(2019)\citenamefont
  {Andrews}, \citenamefont {Jones}, \citenamefont {Reed}, \citenamefont
  {Jones}, \citenamefont {Ha}, \citenamefont {Jura}, \citenamefont {Kerckhoff},
  \citenamefont {Levendorf}, \citenamefont {Meenehan}, \citenamefont {Merkel},
  \citenamefont {Smith}, \citenamefont {Sun}, \citenamefont {Weinstein},
  \citenamefont {Rakher}, \citenamefont {Ladd},\ and\ \citenamefont
  {Borselli}}]{andrews_quantifying_2019}%
  \BibitemOpen
  \bibfield  {author} {\bibinfo {author} {\bibfnamefont {R.~W.}\ \bibnamefont
  {Andrews}}, \bibinfo {author} {\bibfnamefont {C.}~\bibnamefont {Jones}},
  \bibinfo {author} {\bibfnamefont {M.~D.}\ \bibnamefont {Reed}}, \bibinfo
  {author} {\bibfnamefont {A.~M.}\ \bibnamefont {Jones}}, \bibinfo {author}
  {\bibfnamefont {S.~D.}\ \bibnamefont {Ha}}, \bibinfo {author} {\bibfnamefont
  {M.~P.}\ \bibnamefont {Jura}}, \bibinfo {author} {\bibfnamefont
  {J.}~\bibnamefont {Kerckhoff}}, \bibinfo {author} {\bibfnamefont
  {M.}~\bibnamefont {Levendorf}}, \bibinfo {author} {\bibfnamefont
  {S.}~\bibnamefont {Meenehan}}, \bibinfo {author} {\bibfnamefont {S.~T.}\
  \bibnamefont {Merkel}}, \bibinfo {author} {\bibfnamefont {A.}~\bibnamefont
  {Smith}}, \bibinfo {author} {\bibfnamefont {B.}~\bibnamefont {Sun}}, \bibinfo
  {author} {\bibfnamefont {A.~J.}\ \bibnamefont {Weinstein}}, \bibinfo {author}
  {\bibfnamefont {M.~T.}\ \bibnamefont {Rakher}}, \bibinfo {author}
  {\bibfnamefont {T.~D.}\ \bibnamefont {Ladd}},\ and\ \bibinfo {author}
  {\bibfnamefont {M.~G.}\ \bibnamefont {Borselli}},\ }\bibfield  {title}
  {\bibinfo {title} {Quantifying error and leakage in an encoded {Si}/{SiGe}
  triple-dot qubit},\ }\href {https://doi.org/10.1038/s41565-019-0500-4}
  {\bibfield  {journal} {\bibinfo  {journal} {Nat. Nanotechnol.}\ }\textbf
  {\bibinfo {volume} {14}},\ \bibinfo {pages} {747} (\bibinfo {year}
  {2019})}\BibitemShut {NoStop}%
\bibitem [{\citenamefont {{Huang}}\ \emph {et~al.}(2019)\citenamefont
  {{Huang}}, \citenamefont {{Yang}}, \citenamefont {{Chan}}, \citenamefont
  {{Tanttu}}, \citenamefont {{Hensen}}, \citenamefont {{Leon}}, \citenamefont
  {{Fogarty}}, \citenamefont {{Hwang}}, \citenamefont {{Hudson}}, \citenamefont
  {{Itoh}}, \citenamefont {{Morello}}, \citenamefont {{Laucht}},\ and\
  \citenamefont {{Dzurak}}}]{Huang2018}%
  \BibitemOpen
  \bibfield  {author} {\bibinfo {author} {\bibfnamefont {W.}~\bibnamefont
  {{Huang}}}, \bibinfo {author} {\bibfnamefont {C.~H.}\ \bibnamefont {{Yang}}},
  \bibinfo {author} {\bibfnamefont {K.~W.}\ \bibnamefont {{Chan}}}, \bibinfo
  {author} {\bibfnamefont {T.}~\bibnamefont {{Tanttu}}}, \bibinfo {author}
  {\bibfnamefont {B.}~\bibnamefont {{Hensen}}}, \bibinfo {author}
  {\bibfnamefont {R.~C.~C.}\ \bibnamefont {{Leon}}}, \bibinfo {author}
  {\bibfnamefont {M.~A.}\ \bibnamefont {{Fogarty}}}, \bibinfo {author}
  {\bibfnamefont {J.~C.~C.}\ \bibnamefont {{Hwang}}}, \bibinfo {author}
  {\bibfnamefont {F.~E.}\ \bibnamefont {{Hudson}}}, \bibinfo {author}
  {\bibfnamefont {K.~M.}\ \bibnamefont {{Itoh}}}, \bibinfo {author}
  {\bibfnamefont {A.}~\bibnamefont {{Morello}}}, \bibinfo {author}
  {\bibfnamefont {A.}~\bibnamefont {{Laucht}}},\ and\ \bibinfo {author}
  {\bibfnamefont {A.~S.}\ \bibnamefont {{Dzurak}}},\ }\bibfield  {title}
  {\bibinfo {title} {{Fidelity benchmarks for two-qubit gates in silicon}},\
  }\href@noop {} {\bibfield  {journal} {\bibinfo  {journal} {Nature (London)}\
  }\textbf {\bibinfo {volume} {569}},\ \bibinfo {pages} {532} (\bibinfo {year}
  {2019})}\BibitemShut {NoStop}%
\bibitem [{\citenamefont {Xue}\ \emph {et~al.}(2019)\citenamefont {Xue},
  \citenamefont {Watson}, \citenamefont {Helsen}, \citenamefont {Ward},
  \citenamefont {Savage}, \citenamefont {Lagally}, \citenamefont {Coppersmith},
  \citenamefont {Eriksson}, \citenamefont {Wehner},\ and\ \citenamefont
  {Vandersypen}}]{xue_benchmarking_2019}%
  \BibitemOpen
  \bibfield  {author} {\bibinfo {author} {\bibfnamefont {X.}~\bibnamefont
  {Xue}}, \bibinfo {author} {\bibfnamefont {T.}~\bibnamefont {Watson}},
  \bibinfo {author} {\bibfnamefont {J.}~\bibnamefont {Helsen}}, \bibinfo
  {author} {\bibfnamefont {D.}~\bibnamefont {Ward}}, \bibinfo {author}
  {\bibfnamefont {D.}~\bibnamefont {Savage}}, \bibinfo {author} {\bibfnamefont
  {M.}~\bibnamefont {Lagally}}, \bibinfo {author} {\bibfnamefont
  {S.}~\bibnamefont {Coppersmith}}, \bibinfo {author} {\bibfnamefont
  {M.}~\bibnamefont {Eriksson}}, \bibinfo {author} {\bibfnamefont
  {S.}~\bibnamefont {Wehner}},\ and\ \bibinfo {author} {\bibfnamefont
  {L.}~\bibnamefont {Vandersypen}},\ }\bibfield  {title} {\bibinfo {title}
  {Benchmarking {Gate} {Fidelities} in a $\mathrm{Si}/\mathrm{SiGe}$
  {Two}-{Qubit} {Device}},\ }\href {https://doi.org/10.1103/PhysRevX.9.021011}
  {\bibfield  {journal} {\bibinfo  {journal} {Phys. Rev. X}\ }\textbf {\bibinfo
  {volume} {9}},\ \bibinfo {pages} {021011} (\bibinfo {year}
  {2019})}\BibitemShut {NoStop}%
\bibitem [{\citenamefont {Schäffler}(1997)}]{schaffler_high-mobility_1997}%
  \BibitemOpen
  \bibfield  {author} {\bibinfo {author} {\bibfnamefont {F.}~\bibnamefont
  {Schäffler}},\ }\bibfield  {title} {\bibinfo {title} {High-mobility {Si} and
  {Ge} structures},\ }\href {https://doi.org/10.1088/0268-1242/12/12/001}
  {\bibfield  {journal} {\bibinfo  {journal} {Semicond. Sci. Technol.}\
  }\textbf {\bibinfo {volume} {12}},\ \bibinfo {pages} {1515} (\bibinfo {year}
  {1997})}\BibitemShut {NoStop}%
\bibitem [{\citenamefont {Deelman}\ \emph {et~al.}(2016)\citenamefont
  {Deelman}, \citenamefont {Edge},\ and\ \citenamefont
  {Jackson}}]{deelman_metamorphic_2016}%
  \BibitemOpen
  \bibfield  {author} {\bibinfo {author} {\bibfnamefont {P.~W.}\ \bibnamefont
  {Deelman}}, \bibinfo {author} {\bibfnamefont {L.~F.}\ \bibnamefont {Edge}},\
  and\ \bibinfo {author} {\bibfnamefont {C.~A.}\ \bibnamefont {Jackson}},\
  }\bibfield  {title} {\bibinfo {title} {Metamorphic materials for quantum
  computing},\ }\href {https://doi.org/10.1557/mrs.2016.28} {\bibfield
  {journal} {\bibinfo  {journal} {MRS Bull.}\ }\textbf {\bibinfo {volume}
  {41}},\ \bibinfo {pages} {224} (\bibinfo {year} {2016})}\BibitemShut
  {NoStop}%
\bibitem [{\citenamefont {Angus}\ \emph {et~al.}(2007)\citenamefont {Angus},
  \citenamefont {Ferguson}, \citenamefont {Dzurak},\ and\ \citenamefont
  {Clark}}]{angus_gate-defined_2007}%
  \BibitemOpen
  \bibfield  {author} {\bibinfo {author} {\bibfnamefont {S.~J.}\ \bibnamefont
  {Angus}}, \bibinfo {author} {\bibfnamefont {A.~J.}\ \bibnamefont {Ferguson}},
  \bibinfo {author} {\bibfnamefont {A.~S.}\ \bibnamefont {Dzurak}},\ and\
  \bibinfo {author} {\bibfnamefont {R.~G.}\ \bibnamefont {Clark}},\ }\bibfield
  {title} {\bibinfo {title} {Gate-{Defined} {Quantum} {Dots} in {Intrinsic}
  {Silicon}},\ }\href {https://doi.org/10.1021/nl070949k} {\bibfield  {journal}
  {\bibinfo  {journal} {Nano Lett.}\ }\textbf {\bibinfo {volume} {7}},\
  \bibinfo {pages} {2051} (\bibinfo {year} {2007})}\BibitemShut {NoStop}%
\bibitem [{\citenamefont {Borselli}\ \emph {et~al.}(2015)\citenamefont
  {Borselli}, \citenamefont {Eng}, \citenamefont {Ross}, \citenamefont
  {Hazard}, \citenamefont {Holabird}, \citenamefont {Huang}, \citenamefont
  {Kiselev}, \citenamefont {Deelman}, \citenamefont {Warren}, \citenamefont
  {Milosavljevic}, \citenamefont {Schmitz}, \citenamefont {Sokolich},
  \citenamefont {Gyure},\ and\ \citenamefont {Hunter}}]{borselli_undoped_2015}%
  \BibitemOpen
  \bibfield  {author} {\bibinfo {author} {\bibfnamefont {M.~G.}\ \bibnamefont
  {Borselli}}, \bibinfo {author} {\bibfnamefont {K.}~\bibnamefont {Eng}},
  \bibinfo {author} {\bibfnamefont {R.~S.}\ \bibnamefont {Ross}}, \bibinfo
  {author} {\bibfnamefont {T.~M.}\ \bibnamefont {Hazard}}, \bibinfo {author}
  {\bibfnamefont {K.~S.}\ \bibnamefont {Holabird}}, \bibinfo {author}
  {\bibfnamefont {B.}~\bibnamefont {Huang}}, \bibinfo {author} {\bibfnamefont
  {A.~A.}\ \bibnamefont {Kiselev}}, \bibinfo {author} {\bibfnamefont {P.~W.}\
  \bibnamefont {Deelman}}, \bibinfo {author} {\bibfnamefont {L.~D.}\
  \bibnamefont {Warren}}, \bibinfo {author} {\bibfnamefont {I.}~\bibnamefont
  {Milosavljevic}}, \bibinfo {author} {\bibfnamefont {A.~E.}\ \bibnamefont
  {Schmitz}}, \bibinfo {author} {\bibfnamefont {M.}~\bibnamefont {Sokolich}},
  \bibinfo {author} {\bibfnamefont {M.~F.}\ \bibnamefont {Gyure}},\ and\
  \bibinfo {author} {\bibfnamefont {A.~T.}\ \bibnamefont {Hunter}},\ }\bibfield
   {title} {\bibinfo {title} {Undoped accumulation-mode {Si}/{SiGe} quantum
  dots},\ }\href {https://doi.org/10.1088/0957-4484/26/37/375202} {\bibfield
  {journal} {\bibinfo  {journal} {Nanotechnology}\ }\textbf {\bibinfo {volume}
  {26}},\ \bibinfo {pages} {375202} (\bibinfo {year} {2015})}\BibitemShut
  {NoStop}%
\bibitem [{\citenamefont {Zajac}\ \emph {et~al.}(2016)\citenamefont {Zajac},
  \citenamefont {Hazard}, \citenamefont {Mi}, \citenamefont {Nielsen},\ and\
  \citenamefont {Petta}}]{ZajacScalable}%
  \BibitemOpen
  \bibfield  {author} {\bibinfo {author} {\bibfnamefont {D.~M.}\ \bibnamefont
  {Zajac}}, \bibinfo {author} {\bibfnamefont {T.~M.}\ \bibnamefont {Hazard}},
  \bibinfo {author} {\bibfnamefont {X.}~\bibnamefont {Mi}}, \bibinfo {author}
  {\bibfnamefont {E.}~\bibnamefont {Nielsen}},\ and\ \bibinfo {author}
  {\bibfnamefont {J.~R.}\ \bibnamefont {Petta}},\ }\bibfield  {title} {\bibinfo
  {title} {Scalable gate architecture for a one-dimensional array of
  semiconductor spin qubits},\ }\href
  {https://doi.org/10.1103/PhysRevApplied.6.054013} {\bibfield  {journal}
  {\bibinfo  {journal} {Phys. Rev. Appl.}\ }\textbf {\bibinfo {volume} {6}},\
  \bibinfo {pages} {054013} (\bibinfo {year} {2016})}\BibitemShut {NoStop}%
\bibitem [{\citenamefont {Lawrie}\ \emph {et~al.}(2020)\citenamefont {Lawrie},
  \citenamefont {Eenink}, \citenamefont {Hendrickx}, \citenamefont {Boter},
  \citenamefont {Petit}, \citenamefont {Amitonov}, \citenamefont {Lodari},
  \citenamefont {Paquelet~Wuetz}, \citenamefont {Volk}, \citenamefont
  {Philips}, \citenamefont {Droulers}, \citenamefont {Kalhor}, \citenamefont
  {van Riggelen}, \citenamefont {Brousse}, \citenamefont {Sammak},
  \citenamefont {Vandersypen}, \citenamefont {Scappucci},\ and\ \citenamefont
  {Veldhorst}}]{lawrie_quantum_2020}%
  \BibitemOpen
  \bibfield  {author} {\bibinfo {author} {\bibfnamefont {W.~I.~L.}\
  \bibnamefont {Lawrie}}, \bibinfo {author} {\bibfnamefont {H.~G.~J.}\
  \bibnamefont {Eenink}}, \bibinfo {author} {\bibfnamefont {N.~W.}\
  \bibnamefont {Hendrickx}}, \bibinfo {author} {\bibfnamefont {J.~M.}\
  \bibnamefont {Boter}}, \bibinfo {author} {\bibfnamefont {L.}~\bibnamefont
  {Petit}}, \bibinfo {author} {\bibfnamefont {S.~V.}\ \bibnamefont {Amitonov}},
  \bibinfo {author} {\bibfnamefont {M.}~\bibnamefont {Lodari}}, \bibinfo
  {author} {\bibfnamefont {B.}~\bibnamefont {Paquelet~Wuetz}}, \bibinfo
  {author} {\bibfnamefont {C.}~\bibnamefont {Volk}}, \bibinfo {author}
  {\bibfnamefont {S.~G.~J.}\ \bibnamefont {Philips}}, \bibinfo {author}
  {\bibfnamefont {G.}~\bibnamefont {Droulers}}, \bibinfo {author}
  {\bibfnamefont {N.}~\bibnamefont {Kalhor}}, \bibinfo {author} {\bibfnamefont
  {F.}~\bibnamefont {van Riggelen}}, \bibinfo {author} {\bibfnamefont
  {D.}~\bibnamefont {Brousse}}, \bibinfo {author} {\bibfnamefont
  {A.}~\bibnamefont {Sammak}}, \bibinfo {author} {\bibfnamefont {L.~M.~K.}\
  \bibnamefont {Vandersypen}}, \bibinfo {author} {\bibfnamefont
  {G.}~\bibnamefont {Scappucci}},\ and\ \bibinfo {author} {\bibfnamefont
  {M.}~\bibnamefont {Veldhorst}},\ }\bibfield  {title} {\bibinfo {title}
  {Quantum dot arrays in silicon and germanium},\ }\href
  {https://doi.org/10.1063/5.0002013} {\bibfield  {journal} {\bibinfo
  {journal} {Appl. Phys. Lett.}\ }\textbf {\bibinfo {volume} {116}},\ \bibinfo
  {pages} {080501} (\bibinfo {year} {2020})}\BibitemShut {NoStop}%
\bibitem [{\citenamefont {{Mills}}\ \emph {et~al.}(2019)\citenamefont
  {{Mills}}, \citenamefont {{Zajac}}, \citenamefont {{Gullans}}, \citenamefont
  {{Schupp}}, \citenamefont {{Hazard}},\ and\ \citenamefont
  {{Petta}}}]{Mills2018}%
  \BibitemOpen
  \bibfield  {author} {\bibinfo {author} {\bibfnamefont {A.~R.}\ \bibnamefont
  {{Mills}}}, \bibinfo {author} {\bibfnamefont {D.~M.}\ \bibnamefont
  {{Zajac}}}, \bibinfo {author} {\bibfnamefont {M.~J.}\ \bibnamefont
  {{Gullans}}}, \bibinfo {author} {\bibfnamefont {F.~J.}\ \bibnamefont
  {{Schupp}}}, \bibinfo {author} {\bibfnamefont {T.~M.}\ \bibnamefont
  {{Hazard}}},\ and\ \bibinfo {author} {\bibfnamefont {J.~R.}\ \bibnamefont
  {{Petta}}},\ }\bibfield  {title} {\bibinfo {title} {{Shuttling a single
  charge across a one-dimensional array of silicon quantum dots}},\ }\href@noop
  {} {\bibfield  {journal} {\bibinfo  {journal} {Nat. Commun.}\ }\textbf
  {\bibinfo {volume} {10}},\ \bibinfo {pages} {1063} (\bibinfo {year}
  {2019})}\BibitemShut {NoStop}%
\bibitem [{\citenamefont {Zwanenburg}\ \emph {et~al.}(2013)\citenamefont
  {Zwanenburg}, \citenamefont {Dzurak}, \citenamefont {Morello}, \citenamefont
  {Simmons}, \citenamefont {Hollenberg}, \citenamefont {Klimeck}, \citenamefont
  {Rogge}, \citenamefont {Coppersmith},\ and\ \citenamefont
  {Eriksson}}]{zwanenburg2013silicon}%
  \BibitemOpen
  \bibfield  {author} {\bibinfo {author} {\bibfnamefont {F.~A.}\ \bibnamefont
  {Zwanenburg}}, \bibinfo {author} {\bibfnamefont {A.~S.}\ \bibnamefont
  {Dzurak}}, \bibinfo {author} {\bibfnamefont {A.}~\bibnamefont {Morello}},
  \bibinfo {author} {\bibfnamefont {M.~Y.}\ \bibnamefont {Simmons}}, \bibinfo
  {author} {\bibfnamefont {L.~C.~L.}\ \bibnamefont {Hollenberg}}, \bibinfo
  {author} {\bibfnamefont {G.}~\bibnamefont {Klimeck}}, \bibinfo {author}
  {\bibfnamefont {S.}~\bibnamefont {Rogge}}, \bibinfo {author} {\bibfnamefont
  {S.~N.}\ \bibnamefont {Coppersmith}},\ and\ \bibinfo {author} {\bibfnamefont
  {M.~A.}\ \bibnamefont {Eriksson}},\ }\bibfield  {title} {\bibinfo {title}
  {Silicon quantum electronics},\ }\href@noop {} {\bibfield  {journal}
  {\bibinfo  {journal} {Rev. Mod. Phys.}\ }\textbf {\bibinfo {volume} {85}},\
  \bibinfo {pages} {961} (\bibinfo {year} {2013})}\BibitemShut {NoStop}%
\bibitem [{\citenamefont {Goswami}\ \emph {et~al.}(2007)\citenamefont
  {Goswami}, \citenamefont {Slinker}, \citenamefont {Friesen}, \citenamefont
  {McGuire}, \citenamefont {Truitt}, \citenamefont {Tahan}, \citenamefont
  {Klein}, \citenamefont {Chu}, \citenamefont {Mooney}, \citenamefont {van~der
  Weide}, \citenamefont {Joynt}, \citenamefont {Coppersmith},\ and\
  \citenamefont {Eriksson}}]{goswami_controllable_2007}%
  \BibitemOpen
  \bibfield  {author} {\bibinfo {author} {\bibfnamefont {S.}~\bibnamefont
  {Goswami}}, \bibinfo {author} {\bibfnamefont {K.~A.}\ \bibnamefont
  {Slinker}}, \bibinfo {author} {\bibfnamefont {M.}~\bibnamefont {Friesen}},
  \bibinfo {author} {\bibfnamefont {L.~M.}\ \bibnamefont {McGuire}}, \bibinfo
  {author} {\bibfnamefont {J.~L.}\ \bibnamefont {Truitt}}, \bibinfo {author}
  {\bibfnamefont {C.}~\bibnamefont {Tahan}}, \bibinfo {author} {\bibfnamefont
  {L.~J.}\ \bibnamefont {Klein}}, \bibinfo {author} {\bibfnamefont {J.~O.}\
  \bibnamefont {Chu}}, \bibinfo {author} {\bibfnamefont {P.~M.}\ \bibnamefont
  {Mooney}}, \bibinfo {author} {\bibfnamefont {D.~W.}\ \bibnamefont {van~der
  Weide}}, \bibinfo {author} {\bibfnamefont {R.}~\bibnamefont {Joynt}},
  \bibinfo {author} {\bibfnamefont {S.~N.}\ \bibnamefont {Coppersmith}},\ and\
  \bibinfo {author} {\bibfnamefont {M.~A.}\ \bibnamefont {Eriksson}},\
  }\bibfield  {title} {\bibinfo {title} {Controllable valley splitting in
  silicon quantum devices},\ }\href {https://doi.org/10.1038/nphys475}
  {\bibfield  {journal} {\bibinfo  {journal} {Nat. Phys.}\ }\textbf {\bibinfo
  {volume} {3}},\ \bibinfo {pages} {41} (\bibinfo {year} {2007})}\BibitemShut
  {NoStop}%
\bibitem [{\citenamefont {Penthorn}\ \emph {et~al.}(2019)\citenamefont
  {Penthorn}, \citenamefont {Schoenfield}, \citenamefont {Rooney},
  \citenamefont {Edge},\ and\ \citenamefont {Jiang}}]{penthorn_two-axis_2019}%
  \BibitemOpen
  \bibfield  {author} {\bibinfo {author} {\bibfnamefont {N.~E.}\ \bibnamefont
  {Penthorn}}, \bibinfo {author} {\bibfnamefont {J.~S.}\ \bibnamefont
  {Schoenfield}}, \bibinfo {author} {\bibfnamefont {J.~D.}\ \bibnamefont
  {Rooney}}, \bibinfo {author} {\bibfnamefont {L.~F.}\ \bibnamefont {Edge}},\
  and\ \bibinfo {author} {\bibfnamefont {H.}~\bibnamefont {Jiang}},\ }\bibfield
   {title} {\bibinfo {title} {Two-axis quantum control of a fast valley qubit
  in silicon},\ }\href {https://doi.org/10.1038/s41534-019-0212-5} {\bibfield
  {journal} {\bibinfo  {journal} {npj Quantum Inf.}\ }\textbf {\bibinfo
  {volume} {5}},\ \bibinfo {pages} {1} (\bibinfo {year} {2019})}\BibitemShut
  {NoStop}%
\bibitem [{\citenamefont {Friesen}\ \emph {et~al.}(2006)\citenamefont
  {Friesen}, \citenamefont {Eriksson},\ and\ \citenamefont
  {Coppersmith}}]{friesen_magnetic_2006}%
  \BibitemOpen
  \bibfield  {author} {\bibinfo {author} {\bibfnamefont {M.}~\bibnamefont
  {Friesen}}, \bibinfo {author} {\bibfnamefont {M.~A.}\ \bibnamefont
  {Eriksson}},\ and\ \bibinfo {author} {\bibfnamefont {S.~N.}\ \bibnamefont
  {Coppersmith}},\ }\bibfield  {title} {\bibinfo {title} {Magnetic field
  dependence of valley splitting in realistic {Si}/{SiGe} quantum wells},\
  }\href {https://doi.org/10.1063/1.2387975} {\bibfield  {journal} {\bibinfo
  {journal} {Appl. Phys. Lett.}\ }\textbf {\bibinfo {volume} {89}},\ \bibinfo
  {pages} {202106} (\bibinfo {year} {2006})}\BibitemShut {NoStop}%
\bibitem [{\citenamefont {Friesen}\ and\ \citenamefont
  {Coppersmith}(2010)}]{friesen_theory_2010}%
  \BibitemOpen
  \bibfield  {author} {\bibinfo {author} {\bibfnamefont {M.}~\bibnamefont
  {Friesen}}\ and\ \bibinfo {author} {\bibfnamefont {S.~N.}\ \bibnamefont
  {Coppersmith}},\ }\bibfield  {title} {\bibinfo {title} {Theory of
  valley-orbit coupling in a {Si}/{SiGe} quantum dot},\ }\href
  {https://doi.org/10.1103/PhysRevB.81.115324} {\bibfield  {journal} {\bibinfo
  {journal} {Phys. Rev. B}\ }\textbf {\bibinfo {volume} {81}},\ \bibinfo
  {pages} {115324} (\bibinfo {year} {2010})}\BibitemShut {NoStop}%
\bibitem [{\citenamefont {Tariq}\ and\ \citenamefont
  {Hu}(2019)}]{tariq_effects_2019}%
  \BibitemOpen
  \bibfield  {author} {\bibinfo {author} {\bibfnamefont {B.}~\bibnamefont
  {Tariq}}\ and\ \bibinfo {author} {\bibfnamefont {X.}~\bibnamefont {Hu}},\
  }\bibfield  {title} {\bibinfo {title} {Effects of interface steps on the
  valley-orbit coupling in a {Si}/{SiGe} quantum dot},\ }\href
  {https://doi.org/10.1103/PhysRevB.100.125309} {\bibfield  {journal} {\bibinfo
   {journal} {Phys. Rev. B}\ }\textbf {\bibinfo {volume} {100}},\ \bibinfo
  {pages} {125309} (\bibinfo {year} {2019})}\BibitemShut {NoStop}%
\bibitem [{\citenamefont {Elzerman}\ \emph {et~al.}(2004)\citenamefont
  {Elzerman}, \citenamefont {Hanson}, \citenamefont {Van~Beveren},
  \citenamefont {Witkamp}, \citenamefont {Vandersypen},\ and\ \citenamefont
  {Kouwenhoven}}]{elzerman2004single}%
  \BibitemOpen
  \bibfield  {author} {\bibinfo {author} {\bibfnamefont {J.~M.}\ \bibnamefont
  {Elzerman}}, \bibinfo {author} {\bibfnamefont {R.}~\bibnamefont {Hanson}},
  \bibinfo {author} {\bibfnamefont {L.~H.~W.}\ \bibnamefont {Van~Beveren}},
  \bibinfo {author} {\bibfnamefont {B.}~\bibnamefont {Witkamp}}, \bibinfo
  {author} {\bibfnamefont {L.~M.~K.}\ \bibnamefont {Vandersypen}},\ and\
  \bibinfo {author} {\bibfnamefont {L.~P.}\ \bibnamefont {Kouwenhoven}},\
  }\bibfield  {title} {\bibinfo {title} {Single-shot read-out of an individual
  electron spin in a quantum dot},\ }\href@noop {} {\bibfield  {journal}
  {\bibinfo  {journal} {Nature (London)}\ }\textbf {\bibinfo {volume} {430}},\
  \bibinfo {pages} {431} (\bibinfo {year} {2004})}\BibitemShut {NoStop}%
\bibitem [{\citenamefont {Yang}\ \emph {et~al.}(2012)\citenamefont {Yang},
  \citenamefont {Lim}, \citenamefont {Lai}, \citenamefont {Rossi},
  \citenamefont {Morello},\ and\ \citenamefont {Dzurak}}]{yang2012}%
  \BibitemOpen
  \bibfield  {author} {\bibinfo {author} {\bibfnamefont {C.~H.}\ \bibnamefont
  {Yang}}, \bibinfo {author} {\bibfnamefont {W.~H.}\ \bibnamefont {Lim}},
  \bibinfo {author} {\bibfnamefont {N.~S.}\ \bibnamefont {Lai}}, \bibinfo
  {author} {\bibfnamefont {A.}~\bibnamefont {Rossi}}, \bibinfo {author}
  {\bibfnamefont {A.}~\bibnamefont {Morello}},\ and\ \bibinfo {author}
  {\bibfnamefont {A.~S.}\ \bibnamefont {Dzurak}},\ }\bibfield  {title}
  {\bibinfo {title} {Orbital and valley state spectra of a few-electron silicon
  quantum dot},\ }\href@noop {} {\bibfield  {journal} {\bibinfo  {journal}
  {Phys. Rev. B}\ }\textbf {\bibinfo {volume} {86}},\ \bibinfo {pages} {115319}
  (\bibinfo {year} {2012})}\BibitemShut {NoStop}%
\bibitem [{\citenamefont {Petit}\ \emph {et~al.}(2018)\citenamefont {Petit},
  \citenamefont {Boter}, \citenamefont {Eenink}, \citenamefont {Droulers},
  \citenamefont {Tagliaferri}, \citenamefont {Li}, \citenamefont {Franke},
  \citenamefont {Singh}, \citenamefont {Clarke}, \citenamefont {Schouten},
  \citenamefont {Dobrovitski}, \citenamefont {Vandersypen},\ and\ \citenamefont
  {Veldhorst}}]{petit_spin_2018}%
  \BibitemOpen
  \bibfield  {author} {\bibinfo {author} {\bibfnamefont {L.}~\bibnamefont
  {Petit}}, \bibinfo {author} {\bibfnamefont {J.}~\bibnamefont {Boter}},
  \bibinfo {author} {\bibfnamefont {H.}~\bibnamefont {Eenink}}, \bibinfo
  {author} {\bibfnamefont {G.}~\bibnamefont {Droulers}}, \bibinfo {author}
  {\bibfnamefont {M.}~\bibnamefont {Tagliaferri}}, \bibinfo {author}
  {\bibfnamefont {R.}~\bibnamefont {Li}}, \bibinfo {author} {\bibfnamefont
  {D.}~\bibnamefont {Franke}}, \bibinfo {author} {\bibfnamefont
  {K.}~\bibnamefont {Singh}}, \bibinfo {author} {\bibfnamefont
  {J.}~\bibnamefont {Clarke}}, \bibinfo {author} {\bibfnamefont
  {R.}~\bibnamefont {Schouten}}, \bibinfo {author} {\bibfnamefont
  {V.}~\bibnamefont {Dobrovitski}}, \bibinfo {author} {\bibfnamefont
  {L.}~\bibnamefont {Vandersypen}},\ and\ \bibinfo {author} {\bibfnamefont
  {M.}~\bibnamefont {Veldhorst}},\ }\bibfield  {title} {\bibinfo {title} {Spin
  {Lifetime} and {Charge} {Noise} in {Hot} {Silicon} {Quantum} {Dot}
  {Qubits}},\ }\href {https://doi.org/10.1103/PhysRevLett.121.076801}
  {\bibfield  {journal} {\bibinfo  {journal} {Phys. Rev. Lett.}\ }\textbf
  {\bibinfo {volume} {121}},\ \bibinfo {pages} {076801} (\bibinfo {year}
  {2018})}\BibitemShut {NoStop}%
\bibitem [{\citenamefont {Borjans}\ \emph {et~al.}(2019)\citenamefont
  {Borjans}, \citenamefont {Zajac}, \citenamefont {Hazard},\ and\ \citenamefont
  {Petta}}]{Borjans2018}%
  \BibitemOpen
  \bibfield  {author} {\bibinfo {author} {\bibfnamefont {F.}~\bibnamefont
  {Borjans}}, \bibinfo {author} {\bibfnamefont {D.~M.}\ \bibnamefont {Zajac}},
  \bibinfo {author} {\bibfnamefont {T.~M.}\ \bibnamefont {Hazard}},\ and\
  \bibinfo {author} {\bibfnamefont {J.~R.}\ \bibnamefont {Petta}},\ }\bibfield
  {title} {\bibinfo {title} {Single-spin relaxation in a synthetic spin-orbit
  field},\ }\href@noop {} {\bibfield  {journal} {\bibinfo  {journal} {Phys.
  Rev. Appl.}\ }\textbf {\bibinfo {volume} {11}},\ \bibinfo {pages} {044063}
  (\bibinfo {year} {2019})}\BibitemShut {NoStop}%
\bibitem [{\citenamefont {Zhang}\ \emph {et~al.}(2020)\citenamefont {Zhang},
  \citenamefont {Hu}, \citenamefont {Li}, \citenamefont {Jing}, \citenamefont
  {Zhou}, \citenamefont {Ma}, \citenamefont {Ni}, \citenamefont {Luo},
  \citenamefont {Cao}, \citenamefont {Wang}, \citenamefont {Hu}, \citenamefont
  {Jiang}, \citenamefont {Guo},\ and\ \citenamefont {Guo}}]{zhang_giant_2020}%
  \BibitemOpen
  \bibfield  {author} {\bibinfo {author} {\bibfnamefont {X.}~\bibnamefont
  {Zhang}}, \bibinfo {author} {\bibfnamefont {R.-Z.}\ \bibnamefont {Hu}},
  \bibinfo {author} {\bibfnamefont {H.-O.}\ \bibnamefont {Li}}, \bibinfo
  {author} {\bibfnamefont {F.-M.}\ \bibnamefont {Jing}}, \bibinfo {author}
  {\bibfnamefont {Y.}~\bibnamefont {Zhou}}, \bibinfo {author} {\bibfnamefont
  {R.-L.}\ \bibnamefont {Ma}}, \bibinfo {author} {\bibfnamefont
  {M.}~\bibnamefont {Ni}}, \bibinfo {author} {\bibfnamefont {G.}~\bibnamefont
  {Luo}}, \bibinfo {author} {\bibfnamefont {G.}~\bibnamefont {Cao}}, \bibinfo
  {author} {\bibfnamefont {G.-L.}\ \bibnamefont {Wang}}, \bibinfo {author}
  {\bibfnamefont {X.}~\bibnamefont {Hu}}, \bibinfo {author} {\bibfnamefont
  {H.-W.}\ \bibnamefont {Jiang}}, \bibinfo {author} {\bibfnamefont {G.-C.}\
  \bibnamefont {Guo}},\ and\ \bibinfo {author} {\bibfnamefont {G.-P.}\
  \bibnamefont {Guo}},\ }\bibfield  {title} {\bibinfo {title} {Giant
  {Anisotropy} of {Spin} {Relaxation} and {Spin}-{Valley} {Mixing} in a
  {Silicon} {Quantum} {Dot}},\ }\href
  {https://doi.org/10.1103/PhysRevLett.124.257701} {\bibfield  {journal}
  {\bibinfo  {journal} {Phys. Rev. Lett.}\ }\textbf {\bibinfo {volume} {124}},\
  \bibinfo {pages} {257701} (\bibinfo {year} {2020})}\BibitemShut {NoStop}%
\bibitem [{\citenamefont {Friesen}\ \emph {et~al.}(2007)\citenamefont
  {Friesen}, \citenamefont {Chutia}, \citenamefont {Tahan},\ and\ \citenamefont
  {Coppersmith}}]{friesen_valley_2007}%
  \BibitemOpen
  \bibfield  {author} {\bibinfo {author} {\bibfnamefont {M.}~\bibnamefont
  {Friesen}}, \bibinfo {author} {\bibfnamefont {S.}~\bibnamefont {Chutia}},
  \bibinfo {author} {\bibfnamefont {C.}~\bibnamefont {Tahan}},\ and\ \bibinfo
  {author} {\bibfnamefont {S.~N.}\ \bibnamefont {Coppersmith}},\ }\bibfield
  {title} {\bibinfo {title} {Valley splitting theory of
  $\mathrm{Si}\mathrm{Ge}/\mathrm{Si}/\mathrm{Si}\mathrm{Ge}$ quantum wells},\
  }\href {https://doi.org/10.1103/PhysRevB.75.115318} {\bibfield  {journal}
  {\bibinfo  {journal} {Phys. Rev. B}\ }\textbf {\bibinfo {volume} {75}},\
  \bibinfo {pages} {115318} (\bibinfo {year} {2007})}\BibitemShut {NoStop}%
\bibitem [{\citenamefont {Culcer}\ \emph
  {et~al.}(2010{\natexlab{a}})\citenamefont {Culcer}, \citenamefont {Hu},\ and\
  \citenamefont {Das~Sarma}}]{culcer_interface_2010}%
  \BibitemOpen
  \bibfield  {author} {\bibinfo {author} {\bibfnamefont {D.}~\bibnamefont
  {Culcer}}, \bibinfo {author} {\bibfnamefont {X.}~\bibnamefont {Hu}},\ and\
  \bibinfo {author} {\bibfnamefont {S.}~\bibnamefont {Das~Sarma}},\ }\bibfield
  {title} {\bibinfo {title} {Interface roughness, valley-orbit coupling, and
  valley manipulation in quantum dots},\ }\href
  {https://doi.org/10.1103/PhysRevB.82.205315} {\bibfield  {journal} {\bibinfo
  {journal} {Phys. Rev. B}\ }\textbf {\bibinfo {volume} {82}},\ \bibinfo
  {pages} {205315} (\bibinfo {year} {2010}{\natexlab{a}})}\BibitemShut
  {NoStop}%
\bibitem [{\citenamefont {Jiang}\ \emph {et~al.}(2012)\citenamefont {Jiang},
  \citenamefont {Kharche}, \citenamefont {Boykin},\ and\ \citenamefont
  {Klimeck}}]{jiang_effects_2012}%
  \BibitemOpen
  \bibfield  {author} {\bibinfo {author} {\bibfnamefont {Z.}~\bibnamefont
  {Jiang}}, \bibinfo {author} {\bibfnamefont {N.}~\bibnamefont {Kharche}},
  \bibinfo {author} {\bibfnamefont {T.}~\bibnamefont {Boykin}},\ and\ \bibinfo
  {author} {\bibfnamefont {G.}~\bibnamefont {Klimeck}},\ }\bibfield  {title}
  {\bibinfo {title} {Effects of interface disorder on valley splitting in
  {SiGe}/{Si}/{SiGe} quantum wells},\ }\href
  {https://doi.org/10.1063/1.3692174} {\bibfield  {journal} {\bibinfo
  {journal} {Appl. Phys. Lett.}\ }\textbf {\bibinfo {volume} {100}},\ \bibinfo
  {pages} {103502} (\bibinfo {year} {2012})}\BibitemShut {NoStop}%
\bibitem [{\citenamefont {Neyens}\ \emph {et~al.}(2018)\citenamefont {Neyens},
  \citenamefont {Foote}, \citenamefont {Thorgrimsson}, \citenamefont {Knapp},
  \citenamefont {McJunkin}, \citenamefont {Vandersypen}, \citenamefont {Amin},
  \citenamefont {Thomas}, \citenamefont {Clarke}, \citenamefont {Savage},
  \citenamefont {Lagally}, \citenamefont {Friesen}, \citenamefont
  {Coppersmith},\ and\ \citenamefont {Eriksson}}]{neyens_critical_2018}%
  \BibitemOpen
  \bibfield  {author} {\bibinfo {author} {\bibfnamefont {S.~F.}\ \bibnamefont
  {Neyens}}, \bibinfo {author} {\bibfnamefont {R.~H.}\ \bibnamefont {Foote}},
  \bibinfo {author} {\bibfnamefont {B.}~\bibnamefont {Thorgrimsson}}, \bibinfo
  {author} {\bibfnamefont {T.~J.}\ \bibnamefont {Knapp}}, \bibinfo {author}
  {\bibfnamefont {T.}~\bibnamefont {McJunkin}}, \bibinfo {author}
  {\bibfnamefont {L.~M.~K.}\ \bibnamefont {Vandersypen}}, \bibinfo {author}
  {\bibfnamefont {P.}~\bibnamefont {Amin}}, \bibinfo {author} {\bibfnamefont
  {N.~K.}\ \bibnamefont {Thomas}}, \bibinfo {author} {\bibfnamefont {J.~S.}\
  \bibnamefont {Clarke}}, \bibinfo {author} {\bibfnamefont {D.~E.}\
  \bibnamefont {Savage}}, \bibinfo {author} {\bibfnamefont {M.~G.}\
  \bibnamefont {Lagally}}, \bibinfo {author} {\bibfnamefont {M.}~\bibnamefont
  {Friesen}}, \bibinfo {author} {\bibfnamefont {S.~N.}\ \bibnamefont
  {Coppersmith}},\ and\ \bibinfo {author} {\bibfnamefont {M.~A.}\ \bibnamefont
  {Eriksson}},\ }\bibfield  {title} {\bibinfo {title} {The critical role of
  substrate disorder in valley splitting in {Si} quantum wells},\ }\href
  {https://doi.org/10.1063/1.5033447} {\bibfield  {journal} {\bibinfo
  {journal} {Appl. Phys. Lett.}\ }\textbf {\bibinfo {volume} {112}},\ \bibinfo
  {pages} {243107} (\bibinfo {year} {2018})}\BibitemShut {NoStop}%
\bibitem [{\citenamefont {Borselli}\ \emph {et~al.}(2011)\citenamefont
  {Borselli}, \citenamefont {Ross}, \citenamefont {Kiselev}, \citenamefont
  {Croke}, \citenamefont {Holabird}, \citenamefont {Deelman}, \citenamefont
  {Warren}, \citenamefont {Alvarado-Rodriguez}, \citenamefont {Milosavljevic},
  \citenamefont {Ku}, \citenamefont {Wong}, \citenamefont {Schmitz},
  \citenamefont {Sokolich}, \citenamefont {Gyure},\ and\ \citenamefont
  {Hunter}}]{borselli_measurement_2011}%
  \BibitemOpen
  \bibfield  {author} {\bibinfo {author} {\bibfnamefont {M.~G.}\ \bibnamefont
  {Borselli}}, \bibinfo {author} {\bibfnamefont {R.~S.}\ \bibnamefont {Ross}},
  \bibinfo {author} {\bibfnamefont {A.~A.}\ \bibnamefont {Kiselev}}, \bibinfo
  {author} {\bibfnamefont {E.~T.}\ \bibnamefont {Croke}}, \bibinfo {author}
  {\bibfnamefont {K.~S.}\ \bibnamefont {Holabird}}, \bibinfo {author}
  {\bibfnamefont {P.~W.}\ \bibnamefont {Deelman}}, \bibinfo {author}
  {\bibfnamefont {L.~D.}\ \bibnamefont {Warren}}, \bibinfo {author}
  {\bibfnamefont {I.}~\bibnamefont {Alvarado-Rodriguez}}, \bibinfo {author}
  {\bibfnamefont {I.}~\bibnamefont {Milosavljevic}}, \bibinfo {author}
  {\bibfnamefont {F.~C.}\ \bibnamefont {Ku}}, \bibinfo {author} {\bibfnamefont
  {W.~S.}\ \bibnamefont {Wong}}, \bibinfo {author} {\bibfnamefont {A.~E.}\
  \bibnamefont {Schmitz}}, \bibinfo {author} {\bibfnamefont {M.}~\bibnamefont
  {Sokolich}}, \bibinfo {author} {\bibfnamefont {M.~F.}\ \bibnamefont
  {Gyure}},\ and\ \bibinfo {author} {\bibfnamefont {A.~T.}\ \bibnamefont
  {Hunter}},\ }\bibfield  {title} {\bibinfo {title} {Measurement of valley
  splitting in high-symmetry {Si}/{SiGe} quantum dots},\ }\href
  {https://doi.org/10.1063/1.3569717} {\bibfield  {journal} {\bibinfo
  {journal} {Appl. Phys. Lett.}\ }\textbf {\bibinfo {volume} {98}},\ \bibinfo
  {pages} {123118} (\bibinfo {year} {2011})}\BibitemShut {NoStop}%
\bibitem [{\citenamefont {Zajac}\ \emph {et~al.}(2015)\citenamefont {Zajac},
  \citenamefont {Hazard}, \citenamefont {Mi}, \citenamefont {Wang},\ and\
  \citenamefont {Petta}}]{zajac2015reconfigurable}%
  \BibitemOpen
  \bibfield  {author} {\bibinfo {author} {\bibfnamefont {D.~M.}\ \bibnamefont
  {Zajac}}, \bibinfo {author} {\bibfnamefont {T.~M.}\ \bibnamefont {Hazard}},
  \bibinfo {author} {\bibfnamefont {X.}~\bibnamefont {Mi}}, \bibinfo {author}
  {\bibfnamefont {K.}~\bibnamefont {Wang}},\ and\ \bibinfo {author}
  {\bibfnamefont {J.~R.}\ \bibnamefont {Petta}},\ }\bibfield  {title} {\bibinfo
  {title} {A reconfigurable gate architecture for {Si/SiGe} quantum dots},\
  }\href@noop {} {\bibfield  {journal} {\bibinfo  {journal} {Appl. Phys.
  Lett.}\ }\textbf {\bibinfo {volume} {106}},\ \bibinfo {pages} {223507}
  (\bibinfo {year} {2015})}\BibitemShut {NoStop}%
\bibitem [{\citenamefont {Mi}\ \emph {et~al.}(2017{\natexlab{a}})\citenamefont
  {Mi}, \citenamefont {Péterfalvi}, \citenamefont {Burkard},\ and\
  \citenamefont {Petta}}]{mi_high-resolution_2017}%
  \BibitemOpen
  \bibfield  {author} {\bibinfo {author} {\bibfnamefont {X.}~\bibnamefont
  {Mi}}, \bibinfo {author} {\bibfnamefont {C.~G.}\ \bibnamefont {Péterfalvi}},
  \bibinfo {author} {\bibfnamefont {G.}~\bibnamefont {Burkard}},\ and\ \bibinfo
  {author} {\bibfnamefont {J.~R.}\ \bibnamefont {Petta}},\ }\bibfield  {title}
  {\bibinfo {title} {High-{Resolution} {Valley} {Spectroscopy} of {Si}
  {Quantum} {Dots}},\ }\href {https://doi.org/10.1103/PhysRevLett.119.176803}
  {\bibfield  {journal} {\bibinfo  {journal} {Phys. Rev. Lett.}\ }\textbf
  {\bibinfo {volume} {119}},\ \bibinfo {pages} {176803} (\bibinfo {year}
  {2017}{\natexlab{a}})}\BibitemShut {NoStop}%
\bibitem [{\citenamefont {Ferdous}\ \emph {et~al.}(2018)\citenamefont
  {Ferdous}, \citenamefont {Kawakami}, \citenamefont {Scarlino}, \citenamefont
  {Nowak}, \citenamefont {Ward}, \citenamefont {Savage}, \citenamefont
  {Lagally}, \citenamefont {Coppersmith}, \citenamefont {Friesen},
  \citenamefont {Eriksson}, \citenamefont {Vandersypen},\ and\ \citenamefont
  {Rahman}}]{ferdous_valley_2018}%
  \BibitemOpen
  \bibfield  {author} {\bibinfo {author} {\bibfnamefont {R.}~\bibnamefont
  {Ferdous}}, \bibinfo {author} {\bibfnamefont {E.}~\bibnamefont {Kawakami}},
  \bibinfo {author} {\bibfnamefont {P.}~\bibnamefont {Scarlino}}, \bibinfo
  {author} {\bibfnamefont {M.~P.}\ \bibnamefont {Nowak}}, \bibinfo {author}
  {\bibfnamefont {D.~R.}\ \bibnamefont {Ward}}, \bibinfo {author}
  {\bibfnamefont {D.~E.}\ \bibnamefont {Savage}}, \bibinfo {author}
  {\bibfnamefont {M.~G.}\ \bibnamefont {Lagally}}, \bibinfo {author}
  {\bibfnamefont {S.~N.}\ \bibnamefont {Coppersmith}}, \bibinfo {author}
  {\bibfnamefont {M.}~\bibnamefont {Friesen}}, \bibinfo {author} {\bibfnamefont
  {M.~A.}\ \bibnamefont {Eriksson}}, \bibinfo {author} {\bibfnamefont
  {L.~M.~K.}\ \bibnamefont {Vandersypen}},\ and\ \bibinfo {author}
  {\bibfnamefont {R.}~\bibnamefont {Rahman}},\ }\bibfield  {title} {\bibinfo
  {title} {Valley dependent anisotropic spin splitting in silicon quantum
  dots},\ }\href {https://doi.org/10.1038/s41534-018-0075-1} {\bibfield
  {journal} {\bibinfo  {journal} {npj Quantum Inf.}\ }\textbf {\bibinfo
  {volume} {4}},\ \bibinfo {pages} {1} (\bibinfo {year} {2018})}\BibitemShut
  {NoStop}%
\bibitem [{\citenamefont {Hollmann}\ \emph {et~al.}(2020)\citenamefont
  {Hollmann}, \citenamefont {Struck}, \citenamefont {Langrock}, \citenamefont
  {Schmidbauer}, \citenamefont {Schauer}, \citenamefont {Leonhardt},
  \citenamefont {Sawano}, \citenamefont {Riemann}, \citenamefont {Abrosimov},
  \citenamefont {Bougeard},\ and\ \citenamefont
  {Schreiber}}]{hollmann_large_2020}%
  \BibitemOpen
  \bibfield  {author} {\bibinfo {author} {\bibfnamefont {A.}~\bibnamefont
  {Hollmann}}, \bibinfo {author} {\bibfnamefont {T.}~\bibnamefont {Struck}},
  \bibinfo {author} {\bibfnamefont {V.}~\bibnamefont {Langrock}}, \bibinfo
  {author} {\bibfnamefont {A.}~\bibnamefont {Schmidbauer}}, \bibinfo {author}
  {\bibfnamefont {F.}~\bibnamefont {Schauer}}, \bibinfo {author} {\bibfnamefont
  {T.}~\bibnamefont {Leonhardt}}, \bibinfo {author} {\bibfnamefont
  {K.}~\bibnamefont {Sawano}}, \bibinfo {author} {\bibfnamefont
  {H.}~\bibnamefont {Riemann}}, \bibinfo {author} {\bibfnamefont {N.~V.}\
  \bibnamefont {Abrosimov}}, \bibinfo {author} {\bibfnamefont {D.}~\bibnamefont
  {Bougeard}},\ and\ \bibinfo {author} {\bibfnamefont {L.~R.}\ \bibnamefont
  {Schreiber}},\ }\bibfield  {title} {\bibinfo {title} {Large, {Tunable}
  {Valley} {Splitting} and {Single}-{Spin} {Relaxation} {Mechanisms} in a
  $\mathrm{Si}$/$\mathrm{Si}_{x}\mathrm{Ge}_{1\ensuremath{-}x}$ {Quantum}
  {Dot}},\ }\href {https://doi.org/10.1103/PhysRevApplied.13.034068} {\bibfield
   {journal} {\bibinfo  {journal} {Phys. Rev. Appl.}\ }\textbf {\bibinfo
  {volume} {13}},\ \bibinfo {pages} {034068} (\bibinfo {year}
  {2020})}\BibitemShut {NoStop}%
\bibitem [{\citenamefont {Yang}\ \emph {et~al.}(2013)\citenamefont {Yang},
  \citenamefont {Rossi}, \citenamefont {Ruskov}, \citenamefont {Lai},
  \citenamefont {Mohiyaddin}, \citenamefont {Lee}, \citenamefont {Tahan},
  \citenamefont {Klimeck}, \citenamefont {Morello},\ and\ \citenamefont
  {Dzurak}}]{yang2013spin}%
  \BibitemOpen
  \bibfield  {author} {\bibinfo {author} {\bibfnamefont {C.~H.}\ \bibnamefont
  {Yang}}, \bibinfo {author} {\bibfnamefont {A.}~\bibnamefont {Rossi}},
  \bibinfo {author} {\bibfnamefont {R.}~\bibnamefont {Ruskov}}, \bibinfo
  {author} {\bibfnamefont {N.~S.}\ \bibnamefont {Lai}}, \bibinfo {author}
  {\bibfnamefont {F.~A.}\ \bibnamefont {Mohiyaddin}}, \bibinfo {author}
  {\bibfnamefont {S.}~\bibnamefont {Lee}}, \bibinfo {author} {\bibfnamefont
  {C.}~\bibnamefont {Tahan}}, \bibinfo {author} {\bibfnamefont
  {G.}~\bibnamefont {Klimeck}}, \bibinfo {author} {\bibfnamefont
  {A.}~\bibnamefont {Morello}},\ and\ \bibinfo {author} {\bibfnamefont {A.~S.}\
  \bibnamefont {Dzurak}},\ }\bibfield  {title} {\bibinfo {title} {Spin-valley
  lifetimes in a silicon quantum dot with tunable valley splitting},\
  }\href@noop {} {\bibfield  {journal} {\bibinfo  {journal} {Nat. Commun.}\
  }\textbf {\bibinfo {volume} {4}},\ \bibinfo {pages} {2069} (\bibinfo {year}
  {2013})}\BibitemShut {NoStop}%
\bibitem [{\citenamefont {Gamble}\ \emph {et~al.}(2016)\citenamefont {Gamble},
  \citenamefont {Harvey-Collard}, \citenamefont {Jacobson}, \citenamefont
  {Baczewski}, \citenamefont {Nielsen}, \citenamefont {Maurer}, \citenamefont
  {Montaño}, \citenamefont {Rudolph}, \citenamefont {Carroll}, \citenamefont
  {Yang}, \citenamefont {Rossi}, \citenamefont {Dzurak},\ and\ \citenamefont
  {Muller}}]{gamble_valley_2016}%
  \BibitemOpen
  \bibfield  {author} {\bibinfo {author} {\bibfnamefont {J.~K.}\ \bibnamefont
  {Gamble}}, \bibinfo {author} {\bibfnamefont {P.}~\bibnamefont
  {Harvey-Collard}}, \bibinfo {author} {\bibfnamefont {N.~T.}\ \bibnamefont
  {Jacobson}}, \bibinfo {author} {\bibfnamefont {A.~D.}\ \bibnamefont
  {Baczewski}}, \bibinfo {author} {\bibfnamefont {E.}~\bibnamefont {Nielsen}},
  \bibinfo {author} {\bibfnamefont {L.}~\bibnamefont {Maurer}}, \bibinfo
  {author} {\bibfnamefont {I.}~\bibnamefont {Montaño}}, \bibinfo {author}
  {\bibfnamefont {M.}~\bibnamefont {Rudolph}}, \bibinfo {author} {\bibfnamefont
  {M.~S.}\ \bibnamefont {Carroll}}, \bibinfo {author} {\bibfnamefont {C.~H.}\
  \bibnamefont {Yang}}, \bibinfo {author} {\bibfnamefont {A.}~\bibnamefont
  {Rossi}}, \bibinfo {author} {\bibfnamefont {A.~S.}\ \bibnamefont {Dzurak}},\
  and\ \bibinfo {author} {\bibfnamefont {R.~P.}\ \bibnamefont {Muller}},\
  }\bibfield  {title} {\bibinfo {title} {Valley splitting of single-electron
  {Si} {MOS} quantum dots},\ }\href {https://doi.org/10.1063/1.4972514}
  {\bibfield  {journal} {\bibinfo  {journal} {Appl. Phys. Lett.}\ }\textbf
  {\bibinfo {volume} {109}},\ \bibinfo {pages} {253101} (\bibinfo {year}
  {2016})}\BibitemShut {NoStop}%
\bibitem [{\citenamefont {Yang}\ \emph {et~al.}(2020)\citenamefont {Yang},
  \citenamefont {Leon}, \citenamefont {Hwang}, \citenamefont {Saraiva},
  \citenamefont {Tanttu}, \citenamefont {Huang}, \citenamefont
  {Camirand~Lemyre}, \citenamefont {Chan}, \citenamefont {Tan}, \citenamefont
  {Hudson}, \citenamefont {Itoh}, \citenamefont {Morello}, \citenamefont
  {Pioro-Ladrière}, \citenamefont {Laucht},\ and\ \citenamefont
  {Dzurak}}]{yang_operation_2020}%
  \BibitemOpen
  \bibfield  {author} {\bibinfo {author} {\bibfnamefont {C.~H.}\ \bibnamefont
  {Yang}}, \bibinfo {author} {\bibfnamefont {R.~C.~C.}\ \bibnamefont {Leon}},
  \bibinfo {author} {\bibfnamefont {J.~C.~C.}\ \bibnamefont {Hwang}}, \bibinfo
  {author} {\bibfnamefont {A.}~\bibnamefont {Saraiva}}, \bibinfo {author}
  {\bibfnamefont {T.}~\bibnamefont {Tanttu}}, \bibinfo {author} {\bibfnamefont
  {W.}~\bibnamefont {Huang}}, \bibinfo {author} {\bibfnamefont
  {J.}~\bibnamefont {Camirand~Lemyre}}, \bibinfo {author} {\bibfnamefont
  {K.~W.}\ \bibnamefont {Chan}}, \bibinfo {author} {\bibfnamefont {K.~Y.}\
  \bibnamefont {Tan}}, \bibinfo {author} {\bibfnamefont {F.~E.}\ \bibnamefont
  {Hudson}}, \bibinfo {author} {\bibfnamefont {K.~M.}\ \bibnamefont {Itoh}},
  \bibinfo {author} {\bibfnamefont {A.}~\bibnamefont {Morello}}, \bibinfo
  {author} {\bibfnamefont {M.}~\bibnamefont {Pioro-Ladrière}}, \bibinfo
  {author} {\bibfnamefont {A.}~\bibnamefont {Laucht}},\ and\ \bibinfo {author}
  {\bibfnamefont {A.~S.}\ \bibnamefont {Dzurak}},\ }\bibfield  {title}
  {\bibinfo {title} {Operation of a silicon quantum processor unit cell above
  one kelvin},\ }\href {https://doi.org/10.1038/s41586-020-2171-6} {\bibfield
  {journal} {\bibinfo  {journal} {Nature (London)}\ }\textbf {\bibinfo {volume}
  {580}},\ \bibinfo {pages} {350} (\bibinfo {year} {2020})}\BibitemShut
  {NoStop}%
\bibitem [{\citenamefont {Shiau}\ \emph {et~al.}(2007)\citenamefont {Shiau},
  \citenamefont {Chutia},\ and\ \citenamefont {Joynt}}]{shiau_valley_2007}%
  \BibitemOpen
  \bibfield  {author} {\bibinfo {author} {\bibfnamefont {S.-y.}\ \bibnamefont
  {Shiau}}, \bibinfo {author} {\bibfnamefont {S.}~\bibnamefont {Chutia}},\ and\
  \bibinfo {author} {\bibfnamefont {R.}~\bibnamefont {Joynt}},\ }\bibfield
  {title} {\bibinfo {title} {Valley {Kondo} effect in silicon quantum dots},\
  }\href {https://doi.org/10.1103/PhysRevB.75.195345} {\bibfield  {journal}
  {\bibinfo  {journal} {Phys. Rev. B}\ }\textbf {\bibinfo {volume} {75}},\
  \bibinfo {pages} {195345} (\bibinfo {year} {2007})}\BibitemShut {NoStop}%
\bibitem [{\citenamefont {Culcer}\ \emph
  {et~al.}(2010{\natexlab{b}})\citenamefont {Culcer}, \citenamefont
  {Cywiński}, \citenamefont {Li}, \citenamefont {Hu},\ and\ \citenamefont
  {Das~Sarma}}]{culcer_quantum_2010}%
  \BibitemOpen
  \bibfield  {author} {\bibinfo {author} {\bibfnamefont {D.}~\bibnamefont
  {Culcer}}, \bibinfo {author} {\bibfnamefont {L.}~\bibnamefont {Cywiński}},
  \bibinfo {author} {\bibfnamefont {Q.}~\bibnamefont {Li}}, \bibinfo {author}
  {\bibfnamefont {X.}~\bibnamefont {Hu}},\ and\ \bibinfo {author}
  {\bibfnamefont {S.}~\bibnamefont {Das~Sarma}},\ }\bibfield  {title} {\bibinfo
  {title} {Quantum dot spin qubits in silicon: {Multivalley} physics},\ }\href
  {https://doi.org/10.1103/PhysRevB.82.155312} {\bibfield  {journal} {\bibinfo
  {journal} {Phys. Rev. B}\ }\textbf {\bibinfo {volume} {82}},\ \bibinfo
  {pages} {155312} (\bibinfo {year} {2010}{\natexlab{b}})}\BibitemShut
  {NoStop}%
\bibitem [{\citenamefont {Gamble}\ \emph {et~al.}(2013)\citenamefont {Gamble},
  \citenamefont {Eriksson}, \citenamefont {Coppersmith},\ and\ \citenamefont
  {Friesen}}]{gamble_disorder-induced_2013}%
  \BibitemOpen
  \bibfield  {author} {\bibinfo {author} {\bibfnamefont {J.~K.}\ \bibnamefont
  {Gamble}}, \bibinfo {author} {\bibfnamefont {M.~A.}\ \bibnamefont
  {Eriksson}}, \bibinfo {author} {\bibfnamefont {S.~N.}\ \bibnamefont
  {Coppersmith}},\ and\ \bibinfo {author} {\bibfnamefont {M.}~\bibnamefont
  {Friesen}},\ }\bibfield  {title} {\bibinfo {title} {Disorder-induced
  valley-orbit hybrid states in {Si} quantum dots},\ }\href
  {https://doi.org/10.1103/PhysRevB.88.035310} {\bibfield  {journal} {\bibinfo
  {journal} {Phys. Rev. B}\ }\textbf {\bibinfo {volume} {88}},\ \bibinfo
  {pages} {035310} (\bibinfo {year} {2013})}\BibitemShut {NoStop}%
\bibitem [{\citenamefont {Russ}\ \emph {et~al.}(2020)\citenamefont {Russ},
  \citenamefont {Péterfalvi},\ and\ \citenamefont
  {Burkard}}]{russ_theory_2020}%
  \BibitemOpen
  \bibfield  {author} {\bibinfo {author} {\bibfnamefont {M.}~\bibnamefont
  {Russ}}, \bibinfo {author} {\bibfnamefont {C.~G.}\ \bibnamefont
  {Péterfalvi}},\ and\ \bibinfo {author} {\bibfnamefont {G.}~\bibnamefont
  {Burkard}},\ }\bibfield  {title} {\bibinfo {title} {Theory of valley-resolved
  spectroscopy of a {Si} triple quantum dot coupled to a microwave resonator},\
  }\href {https://doi.org/10.1088/1361-648X/ab613f} {\bibfield  {journal}
  {\bibinfo  {journal} {J. Phys.: Condens. Matter}\ }\textbf {\bibinfo {volume}
  {32}},\ \bibinfo {pages} {165301} (\bibinfo {year} {2020})}\BibitemShut
  {NoStop}%
\bibitem [{\citenamefont {Landau}(1932)}]{Zener1932}%
  \BibitemOpen
  \bibfield  {author} {\bibinfo {author} {\bibfnamefont {L.}~\bibnamefont
  {Landau}},\ }\href@noop {} {\bibfield  {journal} {\bibinfo  {journal} {Phys.
  Z. Sowjetunion}\ }\textbf {\bibinfo {volume} {2}},\ \bibinfo {pages} {46}
  (\bibinfo {year} {1932})}\BibitemShut {NoStop}%
\bibitem [{\citenamefont {Zhao}\ and\ \citenamefont
  {Hu}(2018)}]{zhao2018coherent}%
  \BibitemOpen
  \bibfield  {author} {\bibinfo {author} {\bibfnamefont {X.}~\bibnamefont
  {Zhao}}\ and\ \bibinfo {author} {\bibfnamefont {X.}~\bibnamefont {Hu}},\
  }\bibfield  {title} {\bibinfo {title} {Coherent electron transport in silicon
  quantum dots},\ }\href@noop {} {\bibfield  {journal} {\bibinfo  {journal}
  {arXiv:1803.00749}\ } (\bibinfo {year} {2018})}\BibitemShut {NoStop}%
\bibitem [{\citenamefont {Ginzel}\ \emph {et~al.}(2020)\citenamefont {Ginzel},
  \citenamefont {Mills}, \citenamefont {Petta},\ and\ \citenamefont
  {Burkard}}]{ginzel_spin_2020}%
  \BibitemOpen
  \bibfield  {author} {\bibinfo {author} {\bibfnamefont {F.}~\bibnamefont
  {Ginzel}}, \bibinfo {author} {\bibfnamefont {A.~R.}\ \bibnamefont {Mills}},
  \bibinfo {author} {\bibfnamefont {J.~R.}\ \bibnamefont {Petta}},\ and\
  \bibinfo {author} {\bibfnamefont {G.}~\bibnamefont {Burkard}},\ }\bibfield
  {title} {\bibinfo {title} {Spin shuttling in a silicon double quantum dot},\
  }\href {https://doi.org/10.1103/PhysRevB.102.195418} {\bibfield  {journal}
  {\bibinfo  {journal} {Phys. Rev. B}\ }\textbf {\bibinfo {volume} {102}},\
  \bibinfo {pages} {195418} (\bibinfo {year} {2020})}\BibitemShut {NoStop}%
\bibitem [{\citenamefont {Wallraff}\ \emph {et~al.}(2004)\citenamefont
  {Wallraff}, \citenamefont {Schuster}, \citenamefont {Blais}, \citenamefont
  {Frunzio}, \citenamefont {Huang}, \citenamefont {Majer}, \citenamefont
  {Kumar}, \citenamefont {Girvin},\ and\ \citenamefont
  {Schoelkopf}}]{wallraff_strong_2004}%
  \BibitemOpen
  \bibfield  {author} {\bibinfo {author} {\bibfnamefont {A.}~\bibnamefont
  {Wallraff}}, \bibinfo {author} {\bibfnamefont {D.~I.}\ \bibnamefont
  {Schuster}}, \bibinfo {author} {\bibfnamefont {A.}~\bibnamefont {Blais}},
  \bibinfo {author} {\bibfnamefont {L.}~\bibnamefont {Frunzio}}, \bibinfo
  {author} {\bibfnamefont {R.-S.}\ \bibnamefont {Huang}}, \bibinfo {author}
  {\bibfnamefont {J.}~\bibnamefont {Majer}}, \bibinfo {author} {\bibfnamefont
  {S.}~\bibnamefont {Kumar}}, \bibinfo {author} {\bibfnamefont {S.~M.}\
  \bibnamefont {Girvin}},\ and\ \bibinfo {author} {\bibfnamefont {R.~J.}\
  \bibnamefont {Schoelkopf}},\ }\bibfield  {title} {\bibinfo {title} {Strong
  coupling of a single photon to a superconducting qubit using circuit quantum
  electrodynamics},\ }\href {https://doi.org/10.1038/nature02851} {\bibfield
  {journal} {\bibinfo  {journal} {Nature}\ }\textbf {\bibinfo {volume} {431}},\
  \bibinfo {pages} {162} (\bibinfo {year} {2004})}\BibitemShut {NoStop}%
\bibitem [{\citenamefont {Petersson}\ \emph {et~al.}(2012)\citenamefont
  {Petersson}, \citenamefont {McFaul}, \citenamefont {Schroer}, \citenamefont
  {Jung}, \citenamefont {Taylor}, \citenamefont {Houck},\ and\ \citenamefont
  {Petta}}]{petersson2012}%
  \BibitemOpen
  \bibfield  {author} {\bibinfo {author} {\bibfnamefont {K.~D.}\ \bibnamefont
  {Petersson}}, \bibinfo {author} {\bibfnamefont {L.~W.}\ \bibnamefont
  {McFaul}}, \bibinfo {author} {\bibfnamefont {M.~D.}\ \bibnamefont {Schroer}},
  \bibinfo {author} {\bibfnamefont {M.}~\bibnamefont {Jung}}, \bibinfo {author}
  {\bibfnamefont {J.~M.}\ \bibnamefont {Taylor}}, \bibinfo {author}
  {\bibfnamefont {A.~A.}\ \bibnamefont {Houck}},\ and\ \bibinfo {author}
  {\bibfnamefont {J.~R.}\ \bibnamefont {Petta}},\ }\bibfield  {title} {\bibinfo
  {title} {Circuit quantum electrodynamics with a spin qubit},\ }\href
  {http://dx.doi.org/10.1038/nature11559} {\bibfield  {journal} {\bibinfo
  {journal} {Nature (London)}\ }\textbf {\bibinfo {volume} {490}},\ \bibinfo
  {pages} {380} (\bibinfo {year} {2012})}\BibitemShut {NoStop}%
\bibitem [{\citenamefont {Frey}\ \emph {et~al.}(2012)\citenamefont {Frey},
  \citenamefont {Leek}, \citenamefont {Beck}, \citenamefont {Blais},
  \citenamefont {Ihn}, \citenamefont {Ensslin},\ and\ \citenamefont
  {Wallraff}}]{frey_dipole_2012}%
  \BibitemOpen
  \bibfield  {author} {\bibinfo {author} {\bibfnamefont {T.}~\bibnamefont
  {Frey}}, \bibinfo {author} {\bibfnamefont {P.~J.}\ \bibnamefont {Leek}},
  \bibinfo {author} {\bibfnamefont {M.}~\bibnamefont {Beck}}, \bibinfo {author}
  {\bibfnamefont {A.}~\bibnamefont {Blais}}, \bibinfo {author} {\bibfnamefont
  {T.}~\bibnamefont {Ihn}}, \bibinfo {author} {\bibfnamefont {K.}~\bibnamefont
  {Ensslin}},\ and\ \bibinfo {author} {\bibfnamefont {A.}~\bibnamefont
  {Wallraff}},\ }\bibfield  {title} {\bibinfo {title} {Dipole {Coupling} of a
  {Double} {Quantum} {Dot} to a {Microwave} {Resonator}},\ }\href
  {https://doi.org/10.1103/PhysRevLett.108.046807} {\bibfield  {journal}
  {\bibinfo  {journal} {Phys. Rev. Lett.}\ }\textbf {\bibinfo {volume} {108}},\
  \bibinfo {pages} {046807} (\bibinfo {year} {2012})}\BibitemShut {NoStop}%
\bibitem [{\citenamefont {Viennot}\ \emph {et~al.}(2015)\citenamefont
  {Viennot}, \citenamefont {Dartiailh}, \citenamefont {Cottet},\ and\
  \citenamefont {Kontos}}]{viennot_coherent_2015}%
  \BibitemOpen
  \bibfield  {author} {\bibinfo {author} {\bibfnamefont {J.~J.}\ \bibnamefont
  {Viennot}}, \bibinfo {author} {\bibfnamefont {M.~C.}\ \bibnamefont
  {Dartiailh}}, \bibinfo {author} {\bibfnamefont {A.}~\bibnamefont {Cottet}},\
  and\ \bibinfo {author} {\bibfnamefont {T.}~\bibnamefont {Kontos}},\
  }\bibfield  {title} {\bibinfo {title} {Coherent coupling of a single spin to
  microwave cavity photons},\ }\href {https://doi.org/10.1126/science.aaa3786}
  {\bibfield  {journal} {\bibinfo  {journal} {Science}\ }\textbf {\bibinfo
  {volume} {349}},\ \bibinfo {pages} {408} (\bibinfo {year}
  {2015})}\BibitemShut {NoStop}%
\bibitem [{\citenamefont {Stockklauser}\ \emph {et~al.}(2017)\citenamefont
  {Stockklauser}, \citenamefont {Scarlino}, \citenamefont {Koski},
  \citenamefont {Gasparinetti}, \citenamefont {Andersen}, \citenamefont
  {Reichl}, \citenamefont {Wegscheider}, \citenamefont {Ihn}, \citenamefont
  {Ensslin},\ and\ \citenamefont {Wallraff}}]{stock2017}%
  \BibitemOpen
  \bibfield  {author} {\bibinfo {author} {\bibfnamefont {A.}~\bibnamefont
  {Stockklauser}}, \bibinfo {author} {\bibfnamefont {P.}~\bibnamefont
  {Scarlino}}, \bibinfo {author} {\bibfnamefont {J.~V.}\ \bibnamefont {Koski}},
  \bibinfo {author} {\bibfnamefont {S.}~\bibnamefont {Gasparinetti}}, \bibinfo
  {author} {\bibfnamefont {C.~K.}\ \bibnamefont {Andersen}}, \bibinfo {author}
  {\bibfnamefont {C.}~\bibnamefont {Reichl}}, \bibinfo {author} {\bibfnamefont
  {W.}~\bibnamefont {Wegscheider}}, \bibinfo {author} {\bibfnamefont
  {T.}~\bibnamefont {Ihn}}, \bibinfo {author} {\bibfnamefont {K.}~\bibnamefont
  {Ensslin}},\ and\ \bibinfo {author} {\bibfnamefont {A.}~\bibnamefont
  {Wallraff}},\ }\bibfield  {title} {\bibinfo {title} {Strong coupling cavity
  qed with gate-defined double quantum dots enabled by a high impedance
  resonator},\ }\href {https://link.aps.org/doi/10.1103/PhysRevX.7.011030}
  {\bibfield  {journal} {\bibinfo  {journal} {Phys. Rev. X}\ }\textbf {\bibinfo
  {volume} {7}},\ \bibinfo {pages} {011030} (\bibinfo {year}
  {2017})}\BibitemShut {NoStop}%
\bibitem [{\citenamefont {Mi}\ \emph {et~al.}(2017{\natexlab{b}})\citenamefont
  {Mi}, \citenamefont {Cady}, \citenamefont {Zajac}, \citenamefont {Deelman},\
  and\ \citenamefont {Petta}}]{Mi2017}%
  \BibitemOpen
  \bibfield  {author} {\bibinfo {author} {\bibfnamefont {X.}~\bibnamefont
  {Mi}}, \bibinfo {author} {\bibfnamefont {J.~V.}\ \bibnamefont {Cady}},
  \bibinfo {author} {\bibfnamefont {D.~M.}\ \bibnamefont {Zajac}}, \bibinfo
  {author} {\bibfnamefont {P.~W.}\ \bibnamefont {Deelman}},\ and\ \bibinfo
  {author} {\bibfnamefont {J.~R.}\ \bibnamefont {Petta}},\ }\bibfield  {title}
  {\bibinfo {title} {Strong coupling of a single electron in silicon to a
  microwave photon},\ }\href {https://www.ncbi.nlm.nih.gov/pubmed/28008085}
  {\bibfield  {journal} {\bibinfo  {journal} {Science}\ }\textbf {\bibinfo
  {volume} {355}},\ \bibinfo {pages} {156} (\bibinfo {year}
  {2017}{\natexlab{b}})}\BibitemShut {NoStop}%
\bibitem [{\citenamefont {Ando}\ \emph {et~al.}(1982)\citenamefont {Ando},
  \citenamefont {Fowler},\ and\ \citenamefont {Stern}}]{ando1982}%
  \BibitemOpen
  \bibfield  {author} {\bibinfo {author} {\bibfnamefont {T.}~\bibnamefont
  {Ando}}, \bibinfo {author} {\bibfnamefont {A.~B.}\ \bibnamefont {Fowler}},\
  and\ \bibinfo {author} {\bibfnamefont {F.}~\bibnamefont {Stern}},\ }\bibfield
   {title} {\bibinfo {title} {Electronic properties of two-dimensional
  systems},\ }\href@noop {} {\bibfield  {journal} {\bibinfo  {journal} {Rev.
  Mod. Phys.}\ }\textbf {\bibinfo {volume} {54}},\ \bibinfo {pages} {437}
  (\bibinfo {year} {1982})}\BibitemShut {NoStop}%
\bibitem [{\citenamefont {Burkard}\ and\ \citenamefont
  {Petta}(2016)}]{burkard2016dispersive}%
  \BibitemOpen
  \bibfield  {author} {\bibinfo {author} {\bibfnamefont {G.}~\bibnamefont
  {Burkard}}\ and\ \bibinfo {author} {\bibfnamefont {J.~R.}\ \bibnamefont
  {Petta}},\ }\bibfield  {title} {\bibinfo {title} {Dispersive readout of
  valley splittings in cavity-coupled silicon quantum dots},\ }\href@noop {}
  {\bibfield  {journal} {\bibinfo  {journal} {Phys. Rev. B}\ }\textbf {\bibinfo
  {volume} {94}},\ \bibinfo {pages} {195305} (\bibinfo {year}
  {2016})}\BibitemShut {NoStop}%
\bibitem [{\citenamefont {Borjans}\ \emph {et~al.}(2020)\citenamefont
  {Borjans}, \citenamefont {Croot}, \citenamefont {Putz}, \citenamefont {Mi},
  \citenamefont {Quinn}, \citenamefont {Pan}, \citenamefont {Kerckhoff},
  \citenamefont {Pritchett}, \citenamefont {Jackson}, \citenamefont {Edge},
  \citenamefont {Ross}, \citenamefont {Ladd}, \citenamefont {Borselli},
  \citenamefont {Gyure},\ and\ \citenamefont {Petta}}]{Borjans_APL_2020}%
  \BibitemOpen
  \bibfield  {author} {\bibinfo {author} {\bibfnamefont {F.}~\bibnamefont
  {Borjans}}, \bibinfo {author} {\bibfnamefont {X.}~\bibnamefont {Croot}},
  \bibinfo {author} {\bibfnamefont {S.}~\bibnamefont {Putz}}, \bibinfo {author}
  {\bibfnamefont {X.}~\bibnamefont {Mi}}, \bibinfo {author} {\bibfnamefont
  {S.~M.}\ \bibnamefont {Quinn}}, \bibinfo {author} {\bibfnamefont
  {A.}~\bibnamefont {Pan}}, \bibinfo {author} {\bibfnamefont {J.}~\bibnamefont
  {Kerckhoff}}, \bibinfo {author} {\bibfnamefont {E.~J.}\ \bibnamefont
  {Pritchett}}, \bibinfo {author} {\bibfnamefont {C.~A.}\ \bibnamefont
  {Jackson}}, \bibinfo {author} {\bibfnamefont {L.~F.}\ \bibnamefont {Edge}},
  \bibinfo {author} {\bibfnamefont {R.~S.}\ \bibnamefont {Ross}}, \bibinfo
  {author} {\bibfnamefont {T.~D.}\ \bibnamefont {Ladd}}, \bibinfo {author}
  {\bibfnamefont {M.~G.}\ \bibnamefont {Borselli}}, \bibinfo {author}
  {\bibfnamefont {M.~F.}\ \bibnamefont {Gyure}},\ and\ \bibinfo {author}
  {\bibfnamefont {J.~R.}\ \bibnamefont {Petta}},\ }\bibfield  {title} {\bibinfo
  {title} {Split-gate cavity coupler for silicon circuit quantum
  electrodynamics},\ }\href {https://doi.org/10.1063/5.0006442} {\bibfield
  {journal} {\bibinfo  {journal} {Appl. Phys. Lett.}\ }\textbf {\bibinfo
  {volume} {116}},\ \bibinfo {pages} {234001} (\bibinfo {year}
  {2020})}\BibitemShut {NoStop}%
\bibitem [{\citenamefont {Mi}\ \emph {et~al.}(2017{\natexlab{c}})\citenamefont
  {Mi}, \citenamefont {Cady}, \citenamefont {Zajac}, \citenamefont {Stehlik},
  \citenamefont {Edge},\ and\ \citenamefont {Petta}}]{MiAPL2017}%
  \BibitemOpen
  \bibfield  {author} {\bibinfo {author} {\bibfnamefont {X.}~\bibnamefont
  {Mi}}, \bibinfo {author} {\bibfnamefont {J.~V.}\ \bibnamefont {Cady}},
  \bibinfo {author} {\bibfnamefont {D.~M.}\ \bibnamefont {Zajac}}, \bibinfo
  {author} {\bibfnamefont {J.}~\bibnamefont {Stehlik}}, \bibinfo {author}
  {\bibfnamefont {L.~F.}\ \bibnamefont {Edge}},\ and\ \bibinfo {author}
  {\bibfnamefont {J.~R.}\ \bibnamefont {Petta}},\ }\bibfield  {title} {\bibinfo
  {title} {Circuit quantum electrodynamics architecture for gate-defined
  quantum dots in silicon},\ }\href@noop {} {\bibfield  {journal} {\bibinfo
  {journal} {Appl. Phys. Lett.}\ }\textbf {\bibinfo {volume} {110}},\ \bibinfo
  {pages} {043502} (\bibinfo {year} {2017}{\natexlab{c}})}\BibitemShut
  {NoStop}%
\bibitem [{\citenamefont {Corrigan}\ \emph {et~al.}(2020)\citenamefont
  {Corrigan}, \citenamefont {Dodson}, \citenamefont {Ercan}, \citenamefont
  {Abadillo-Uriel}, \citenamefont {Thorgrimsson}, \citenamefont {Knapp},
  \citenamefont {Holman}, \citenamefont {McJunkin}, \citenamefont {Neyens},
  \citenamefont {MacQuarrie}, \citenamefont {Foote}, \citenamefont {Edge},
  \citenamefont {Friesen}, \citenamefont {Coppersmith},\ and\ \citenamefont
  {Eriksson}}]{corrigan_coherent_2020}%
  \BibitemOpen
  \bibfield  {author} {\bibinfo {author} {\bibfnamefont {J.}~\bibnamefont
  {Corrigan}}, \bibinfo {author} {\bibfnamefont {J.~P.}\ \bibnamefont
  {Dodson}}, \bibinfo {author} {\bibfnamefont {H.~E.}\ \bibnamefont {Ercan}},
  \bibinfo {author} {\bibfnamefont {J.~C.}\ \bibnamefont {Abadillo-Uriel}},
  \bibinfo {author} {\bibfnamefont {B.}~\bibnamefont {Thorgrimsson}}, \bibinfo
  {author} {\bibfnamefont {T.~J.}\ \bibnamefont {Knapp}}, \bibinfo {author}
  {\bibfnamefont {N.}~\bibnamefont {Holman}}, \bibinfo {author} {\bibfnamefont
  {T.}~\bibnamefont {McJunkin}}, \bibinfo {author} {\bibfnamefont {S.~F.}\
  \bibnamefont {Neyens}}, \bibinfo {author} {\bibfnamefont {E.~R.}\
  \bibnamefont {MacQuarrie}}, \bibinfo {author} {\bibfnamefont {R.~H.}\
  \bibnamefont {Foote}}, \bibinfo {author} {\bibfnamefont {L.~F.}\ \bibnamefont
  {Edge}}, \bibinfo {author} {\bibfnamefont {M.}~\bibnamefont {Friesen}},
  \bibinfo {author} {\bibfnamefont {S.~N.}\ \bibnamefont {Coppersmith}},\ and\
  \bibinfo {author} {\bibfnamefont {M.~A.}\ \bibnamefont {Eriksson}},\
  }\bibfield  {title} {\bibinfo {title} {Coherent control and spectroscopy of a
  semiconductor quantum dot {Wigner} molecule},\ }\href
  {http://arxiv.org/abs/2009.13572} {\bibfield  {journal} {\bibinfo  {journal}
  {arXiv:2009.13572}\ } (\bibinfo {year} {2020})}\BibitemShut {NoStop}%
\bibitem [{\citenamefont {Benito}\ \emph {et~al.}(2017)\citenamefont {Benito},
  \citenamefont {Mi}, \citenamefont {Taylor}, \citenamefont {Petta},\ and\
  \citenamefont {Burkard}}]{benito2017input}%
  \BibitemOpen
  \bibfield  {author} {\bibinfo {author} {\bibfnamefont {M.}~\bibnamefont
  {Benito}}, \bibinfo {author} {\bibfnamefont {X.}~\bibnamefont {Mi}}, \bibinfo
  {author} {\bibfnamefont {J.~M.}\ \bibnamefont {Taylor}}, \bibinfo {author}
  {\bibfnamefont {J.~R.}\ \bibnamefont {Petta}},\ and\ \bibinfo {author}
  {\bibfnamefont {G.}~\bibnamefont {Burkard}},\ }\bibfield  {title} {\bibinfo
  {title} {Input-output theory for spin-photon coupling in si double quantum
  dots},\ }\href@noop {} {\bibfield  {journal} {\bibinfo  {journal} {Phys. Rev.
  B}\ }\textbf {\bibinfo {volume} {96}},\ \bibinfo {pages} {235434} (\bibinfo
  {year} {2017})}\BibitemShut {NoStop}%
\bibitem [{\citenamefont {Maune}\ \emph {et~al.}(2012)\citenamefont {Maune},
  \citenamefont {Borselli}, \citenamefont {Huang}, \citenamefont {Ladd},
  \citenamefont {Deelman}, \citenamefont {Holabird}, \citenamefont {Kiselev},
  \citenamefont {Alvarado-Rodriguez}, \citenamefont {Ross}, \citenamefont
  {Schmitz}, \citenamefont {Sokolich}, \citenamefont {Watson}, \citenamefont
  {Gyure},\ and\ \citenamefont {Hunter}}]{maune2012}%
  \BibitemOpen
  \bibfield  {author} {\bibinfo {author} {\bibfnamefont {B.~M.}\ \bibnamefont
  {Maune}}, \bibinfo {author} {\bibfnamefont {M.~G.}\ \bibnamefont {Borselli}},
  \bibinfo {author} {\bibfnamefont {B.}~\bibnamefont {Huang}}, \bibinfo
  {author} {\bibfnamefont {T.~D.}\ \bibnamefont {Ladd}}, \bibinfo {author}
  {\bibfnamefont {P.~W.}\ \bibnamefont {Deelman}}, \bibinfo {author}
  {\bibfnamefont {K.~S.}\ \bibnamefont {Holabird}}, \bibinfo {author}
  {\bibfnamefont {A.~A.}\ \bibnamefont {Kiselev}}, \bibinfo {author}
  {\bibfnamefont {I.}~\bibnamefont {Alvarado-Rodriguez}}, \bibinfo {author}
  {\bibfnamefont {R.~S.}\ \bibnamefont {Ross}}, \bibinfo {author}
  {\bibfnamefont {A.~E.}\ \bibnamefont {Schmitz}}, \bibinfo {author}
  {\bibfnamefont {M.}~\bibnamefont {Sokolich}}, \bibinfo {author}
  {\bibfnamefont {C.~A.}\ \bibnamefont {Watson}}, \bibinfo {author}
  {\bibfnamefont {M.~F.}\ \bibnamefont {Gyure}},\ and\ \bibinfo {author}
  {\bibfnamefont {A.~T.}\ \bibnamefont {Hunter}},\ }\bibfield  {title}
  {\bibinfo {title} {Coherent singlet-triplet oscillations in a silicon-based
  double quantum dot},\ }\href@noop {} {\bibfield  {journal} {\bibinfo
  {journal} {Nature (London)}\ }\textbf {\bibinfo {volume} {481}},\ \bibinfo
  {pages} {344} (\bibinfo {year} {2012})}\BibitemShut {NoStop}%
\end{thebibliography}%

\end{document}